\documentclass[letterpaper]{article} 
\usepackage{aaai2026}  
\usepackage{times}  
\usepackage{helvet}  
\usepackage{courier}  
\usepackage[hyphens]{url}  
\usepackage{graphicx} 
\usepackage{natbib}  
\usepackage{caption} 
\usepackage{algorithm}
\usepackage{algorithmic}
\usepackage{listings}
\usepackage{tabularx}
\usepackage{multirow}
\usepackage{utfsym}
\usepackage{graphicx}
\usepackage{float}
\usepackage[algo2e,ruled,vlined]{algorithm2e}
\usepackage{amsmath}
\usepackage{amssymb}
\usepackage{bm}
\usepackage{booktabs}
\usepackage{amsfonts}
\newcommand{\eg}{e.g.,\ }

\definecolor{codegray}{gray}{0.9}
\usepackage{newfloat}
\usepackage{listings}
\DeclareCaptionStyle{ruled}{labelfont=normalfont,labelsep=colon,strut=off} 
\floatstyle{ruled}
\newfloat{listing}{tb}{lst}{}
\floatname{listing}{Listing}
\nocopyright
\title{OptMark: Robust Multi-bit Diffusion Watermarking via \\ Inference Time Optimization}
\author{
    Jiazheng Xing \textsuperscript{\rm 1, 2}\equalcontrib,
    Hai Ci \textsuperscript{\rm 2}\equalcontrib,
    Hongbin Xu \textsuperscript{\rm 2},
    Hangjie Yuan \textsuperscript{\rm 1},
    Yong Liu \textsuperscript{\rm 1}\thanks{Corresponding author.},
    Mike Zheng Shou \textsuperscript{\rm 2}
}
\affiliations{
    \textsuperscript{\rm 1}Zhejiang University,
    \textsuperscript{\rm 2}
Show Lab, National University of Singapore \\
    jiazhengxing@zju.edu.cn, yongliu@iipc.zju.edu.cn  \\
    \{cihai03, mike.zheng.shou\}@gmail.com \\
     Project Page: \url{https://jiazheng-xing.github.io/optmark-home}
}

\usepackage{bibentry}

\begin{document}

\maketitle

\begin{abstract}
Watermarking diffusion-generated images is crucial for copyright protection and user tracking.
However, current diffusion watermarking methods face significant limitations: zero-bit watermarking systems lack the capacity for large-scale user tracking, while multi-bit methods are highly sensitive to certain image transformations or generative attacks, resulting in a lack of comprehensive robustness.
In this paper, we propose \textbf{OptMark}, an optimization-based approach that embeds a robust multi-bit watermark into the intermediate latents of the diffusion denoising process. OptMark strategically inserts a structural watermark early to resist generative attacks and a detail watermark late to withstand image transformations, with tailored regularization terms to preserve image quality and ensure imperceptibility.
To address the challenge of memory consumption growing linearly with the number of denoising steps during optimization, OptMark incorporates adjoint gradient methods, reducing memory usage from $O(N)$ to $O(1)$. Experimental results demonstrate that OptMark achieves invisible multi-bit watermarking while ensuring robust resilience against valuemetric transformations, geometric transformations, editing, and regeneration attacks.

\end{abstract}    
\section{Introduction}
\label{sec:introduction}

In the AIGC era, diffusion models~\cite{ho2020denoising, song2020denoising, rombach2022high} have become a cornerstone of digital content creation, enabling the generation of hyper-realistic images. This advancement revolutionizes visual content production while raising critical intellectual property and content safety challenges in the digital age. As a crucial copyright protection technology, invisible watermarking enables AIGC service providers to embed imperceptible identifiers into generated content, facilitating traceability and ownership verification. This paper explores multi-bit invisible watermarking for diffusion-generated content, focusing on copyright protection and traceability. 


{Current watermarking approaches fall into two camps: pixel-level and semantic-level.}
Pixel-level watermarking methods, such as HiDDeN~\cite{zhu2018hidden}, SSL~\cite{fernandez2022watermarking}, WAM~\cite{sander2024watermark}, and Stable Signature~\cite{fernandez2023stable}, embed watermarks directly at the pixel level. While these methods are straightforward to implement, they exhibit limited robustness against regeneration attacks~\cite{zhao2023invisible}. 
Semantic-level watermarking methods usually embed watermarks during the image generation process and alter the semantic layout of the generated images. A typical approach is to embed handcrafted watermark patterns in the diffusion noise. Compared with pixel-level methods, these approaches are more robust to regeneration attacks, yet they remain vulnerable to certain image transformations and often lack sufficient capacity to embed more bits. Specifically,
Tree-Ring~\cite{wen2023tree} is susceptible to cropping and scaling, while Gaussian Shading~\cite{yang2024gaussian} is vulnerable to geometric attacks that disrupt the order of patches, such as horizontal flipping. Furthermore, methods such as RingID~\cite{ci2024ringid} and WIND~\cite{arabi2024hidden} lack sufficient capacity to embed adequate watermark bits, limiting their scalability. 
Overall, significant challenges remain in balancing robustness and capacity in existing approaches.


\begin{figure*}[t]
    \centering
    \includegraphics[width=0.845\linewidth]{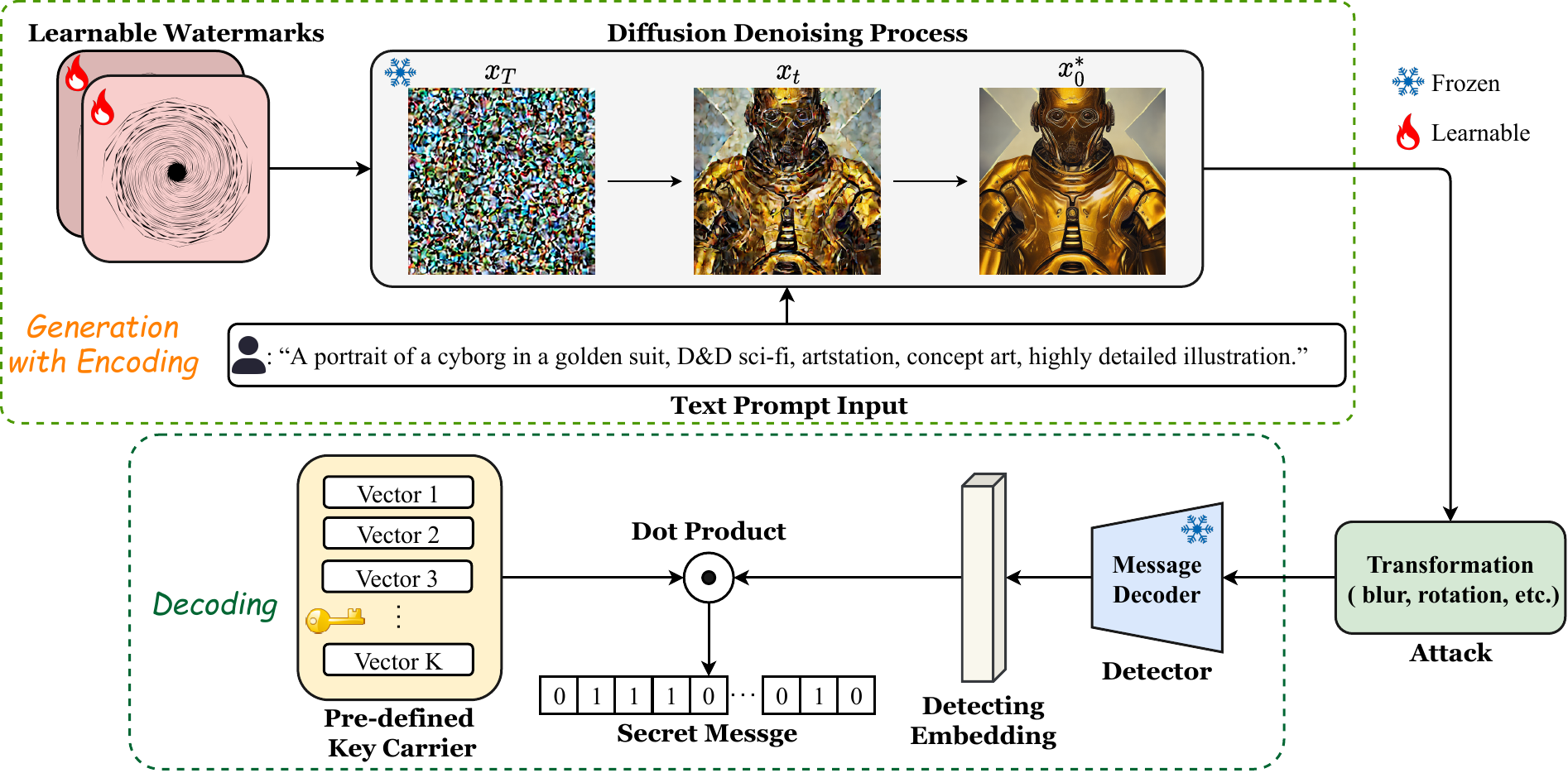}
    \vskip -0.1in
    \caption{{Pipeline of our end-to-end optimized \textbf{OptMark}.} The robust watermark is embedded into the diffusion latent space during the generation process through inference time optimization. In the Decoding phase, the watermark embedding is extracted using a pre-trained message decoder, and the secret message is retrieved by comparing the decoded watermark embedding against a predefined key carrier.}
    \label{fig:main}
    \vskip -0.1in
\end{figure*}

In this paper, we propose \textbf{OptMark}, a novel semantic-level multi-bit watermarking approach that ensures ample capacity while achieving comprehensive robustness against four common types of attacks: valuemetric, geometric, editing, and regeneration, as shown in Fig.~\ref{fig:main}. 
To achieve this, OptMark optimizes the watermarks in an end-to-end manner during the diffusion inference process.
Unlike prior works that rely on handcrafted watermark patterns~\cite{wen2023tree,ci2024ringid,yang2024gaussian}, our approach offers two key advantages through end-to-end learning:
1) \emph{Enhanced robustness}: By seamlessly integrating with diverse training-time image augmentations, OptMark improves resilience against a wide range of attacks, whereas manually designed watermarks struggle to cover all possible scenarios.
2) \emph{Greater flexibility}: 
End-to-end optimization allows for the efficient embedding of a larger number of bits, as the process is fully automated.

To establish this end-to-end optimization framework with comprehensive robustness, high image quality, and low GPU memory overhead, we introduce three key designs: 1) \textbf{Comprehensive Robustness}: We adopt a dual watermarking mechanism, optimizing a structure watermark in the initial diffusion noise to resist generative attacks and a detail watermark 
{in one late denoising step} to counter image transformations. 
2) \textbf{Minimal Impact on Image Quality}: We develop specialized embedding strategies and constraints to regulate the shape and statistical properties of the learned watermarks, ensuring high image quality and imperceptibility. 3) \textbf{Efficient GPU Memory Usage}: To reduce GPU memory overhead, we introduce the adjoint method for computing gradients on learnable watermarks, lowering memory consumption from $O(N)$ to $O(1)$. {Extensive experiments demonstrate that our method significantly outperforms state-of-the-art approaches in robustness, with sufficient bit capacity and high generated image quality.}






\section{Related Work}
\label{sec:related_work}
\subsection{Pixel-Level Watermark}
Pixel-level watermarking typically embeds invisible watermarks directly into the image pixel domain. Mainstream approaches can be categorized into two types: optimization-based methods and encoder-decoder methods. Representative optimization-based approaches, such as FNNS~\cite{kishore2021fixed} and SSL~\cite{fernandez2022watermarking}, iteratively optimize a small perturbation on the cover image so that the image features extracted by a pre-trained model can reliably recover the target watermark bits. In contrast, encoder-decoder methods~\cite{zhu2018hidden,tancik2020stegastamp,fernandez2023stable,ci2024wmadapter,sander2024watermark} train watermark encoders and decoders on a large set of images with different watermark bit sequences, enabling on-the-fly embedding of watermark bits into images. While pixel-level watermarking is imperceptible to the human eye, it has been shown to be inherently vulnerable to regeneration attacks~\cite{zhao2023invisible}.

\subsection{Semantic-Level Watermark}
Semantic-level watermark approaches embed watermarks during the diffusion generation process, altering the semantic content and layout of the generated image, and improving robustness against regeneration attacks. Some methods train diffusion plugins~\cite{feng2024aqualora,min2024watermark} for semantic watermarking, but they require expensive training and struggle to achieve optimal robustness. While Tree-Ring~\cite{wen2023tree} pioneered another direction by injecting a handcrafted tree-ring pattern into the initial diffusion noise as a zero-bit watermark. Subsequent works~\cite{ci2024ringid,yang2024gaussian,zodiac,huang2024robin,gunn2024undetectable} have improved its robustness or imperceptibility. However, they either remain vulnerable to geometric attacks~\cite{yang2024gaussian} or lack the capacity to embed sufficient multi-bit information~\cite{ci2024ringid,zodiac,huang2024robin}.
Our proposed method, OptMark, belongs to the semantic-level watermarking. It is the first approach to achieve both sufficient multi-bit capacity and comprehensive robustness against common image transformations and generative attacks.
\section{Method}
\label{sec:method}

\subsection{Preliminary}
\noindent \textbf{Diffusion Models.} 
{Diffusion models~\cite{ho2020denoising, song2020denoising} progressively convert standard Gaussian noise $x_T \sim \mathcal{N}(0,\mathbf I)$ into samples from the true data distribution $x_0 \sim q(x)$ over $T$ reverse (denoising) steps. The {forward (noising) process} is defined as:
} 

\begin{equation}
    q(x_t|x_{t-1})=\mathcal{N}(x_t; \sqrt[]{1-\beta_t}x_{t-1},\beta _t\mathbf{I}),
\end{equation}
where $\left \{  \beta _{t}\right \} _{t=1}^{T} \in \left( 0,1 \right)$  is the scheduled variance, {and $x_t$ can be sampled directly from $x_0$ as:}

\begin{equation}
    x_t=\sqrt[]{\bar{\alpha }_t }x_0+\sqrt[]{1-\bar{\alpha }_t }\epsilon,
\end{equation}
where $\bar{\alpha }_t= {\textstyle \prod_{i=0}^{t}(1-\beta_t)}$ and $\epsilon \sim \mathcal{N}(0,\mathbf{I} ) $. Subsequently, a network $\epsilon_{\theta}$ is trained to {predict the added noise at each step}, with the following objective:
\begin{equation}  
\mathbb{E} _{x_0, t\sim \texttt{Uniform}(1,T), \epsilon \in \mathcal{N} \left( 0,\textbf{I} \right)}\left[ \left\| \epsilon -\epsilon _{\theta }\left( x_t,{t},\psi(p)\right) \right\| _{2}^{2}\right],
\end{equation}
where $x_{t}$ represents the noisy latent at timesteps $t$ and $\psi(p)$ denotes the embedding of the text input prompt $p$. The  {reverse (generation)} process can be written as:
\begin{align}
\begin{split}
\label{eq:ddim_sampling}
    x_{t-1} = &\sqrt{\alpha_{t-1}}\left(\frac{x_t-\sqrt{1-\alpha_t}\epsilon_{\theta}\left(x_t\right)}{\sqrt{\alpha_t}}\right) \\ &+ \sqrt{1-\alpha_{t-1}-\sigma_t^2}\cdot\epsilon_{\theta}\left(x_t\right) + \sigma_t\epsilon_t.
\end{split}
\end{align}
When $\sigma_t=0$, it is a DDIM sampler~\cite{song2020denoising}. When $\sigma_t=\sqrt{\left(1-\alpha_{t-1}\right)/\left(1-\alpha_t\right)}\sqrt{1-\alpha_t/\alpha_{t-1}}$, it is a DDPM sampler~\cite{ho2020denoising}.

\noindent \textbf{Background and Task Definition.}
{In the multi‑bit watermarking scenario for diffusion models, OptMark embeds a $k$-bit invisible watermark message $m$ into the generation process to produce a watermarked image $x_0^*$.} {When these images are disseminated online, they may undergo various attacks $\mathcal{T}$.} {For copyright verification or user identification, the model owner decodes the potentially distorted image $\mathcal{T}(x_0^*)$ to recover $\hat{m}$ and compares it to the original watermark $m$.}

\subsection{Overview}
{Figure~\ref{fig:main} illustrates the OptMark’s end-to-end pipeline, which comprises two stages: \emph{Watermark Encoding} and \emph{Decoding}. In the \emph{Watermark Encoding} stage, learnable watermark vectors are injected into the diffusion latents during inference to produce a watermarked image $x_0^*$. An inference‑time optimization strategy balances watermark robustness against visual fidelity. In the \emph{Decoding} stage, we employ a pre‑trained, self‑supervised image encoder~\cite{caron2021emerging} as the message decoder to extract the embedded watermark representation from versions of  $x_0^*$ subjected to attacks $\mathcal T$. Finally, the $k$-bit message is recovered by computing the dot product between this representation and a pre-defined set of $k$ carrier vectors.}

\subsection{Dual-Watermark for Diffusion Models}
\paragraph{Watermark Encoding}
{Compared with recent pixel‑level watermarking methods~\cite{kishore2021fixed, fernandez2022watermarking}, which exhibit poor robustness against regeneration attacks~\cite{zhao2023invisible}, OptMark embeds messages directly into the diffusion denoising process and thus achieves significantly higher resistance to these attacks. 
The diffusion model’s denoising trajectory can be divided into two stages: \emph{structure formulation} and \emph{detail refinement}. We therefore propose injecting different watermarks at each stage, with each targeting a distinct semantic level, to enhance robustness against a wide range of attacks.} However, since imprinting the watermark into the denoising process is an increasing entropy reaction, excessive introduction of the watermark can negatively impact the quality of image generation. {To balance the watermark robustness and image quality, OptMark inserts exactly one watermark per stage: a \emph{structure watermark} during the first stage, injected into high-level semantic features to leave a persistent mark that is difficult to erase through generative attacks; and a \emph{detail watermark} during the second stage embedded at a finer, near-pixel level to withstand geometric and volumetric attacks while accelerating convergence.}


\begin{figure*} [ht!]
	\centering
	\includegraphics[width=0.84\linewidth]{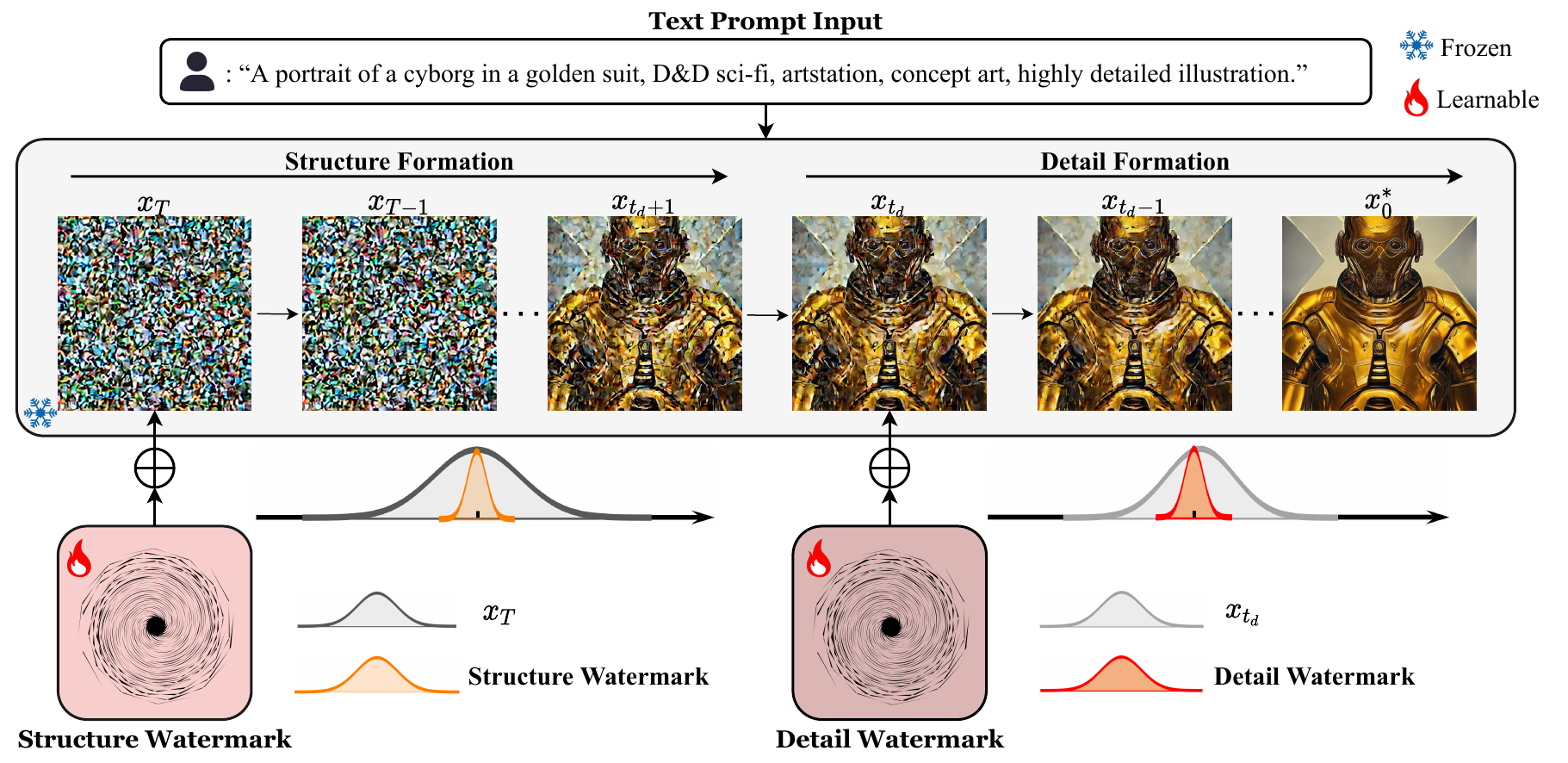}
        \vskip -0.1in
        \caption{{OptMark’s imprinting process consists of two sequential stages: first, a structure watermark is injected into the initial latent state of generation; then, a detail watermark is embedded at an intermediate timestep. These complementary watermarks work in concert to maximize overall robustness.
}}
	\label{fig:message_embed}
    \vskip -0.1in
\end{figure*}

We consider a standard diffusion framework {using the DDIM sampler~\cite{song2020denoising}.} Fig.~\ref{fig:message_embed} depicts the watermark embedding process in OptMark. Given standard Gaussian initial noise $x_T\sim \mathcal{N}\left( 0,\mathbf{I} \right)$, the model predicts the noise $\epsilon _{\theta }$ at each denoising timestep $t$ via:

\begin{equation}
    \hat{\epsilon}_t=
\begin{cases}
\epsilon _\theta \left(\mathcal{F}_{s}\left(x_t,w_s\right),t,\psi\left(p\right) \right)
 & \text{ if }  t = t_{s},  \\
\epsilon _\theta \left(\mathcal{F}_{d}\left(x_t,w_d\right),t,\psi(p) \right) & \text{ if } t = t_{d}, \\
\epsilon _\theta (x_t,t,\psi(p)) & \text{otherwise},
\end{cases}
\end{equation}
where $w_s$ and $w_d$ represent the structure and detail watermark, respectively, both initialized with a Gaussian distribution. {$\mathcal{F}_{s}$ and $\mathcal{F}_{d}$ specify the corresponding watermark‑embedding operator.}  

\begin{figure} [t!]
	\centering
	\includegraphics[width=0.95\linewidth]{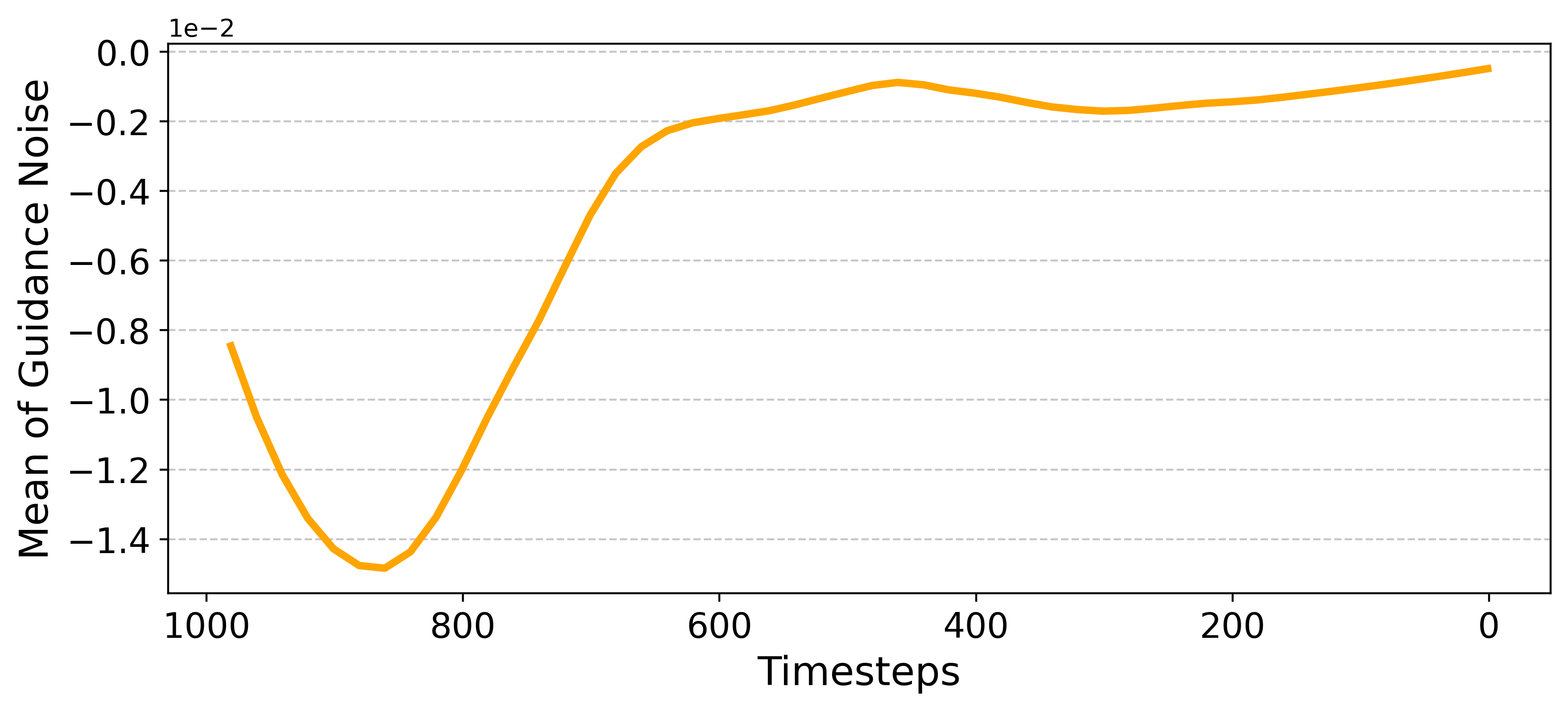}
        \vskip -0.1in
	\caption{Predicted Guidance Noise during generation.}
	\label{fig:noise_line}
    \vskip -0.1in
\end{figure}

\paragraph{Choices of Watermark Position}
{We inject the structure watermark $w_s$ at the initial timestep  $t_{s}=T$ for two reasons:
(i) injecting at initialization enhances robustness against generative attacks; and
(ii) the latents $x_T$  follow the standard normal distribution $\mathcal{N}\left( 0,\mathbf I \right)$, which serves as a reference to constrain the post‑embedding distribution and thus minimize any degradation in generation quality.}
For the detail watermark $w_d$, we need to select an appropriate timestep $t_{d}$ after the semantic generation process, ensuring that the introduction of $w_d$ does not distort the semantics of the generated image. At the same time, this step should not be too close to the pixel level, as pixel-level watermarks are more vulnerable to regeneration attacks and prone to introducing visible artifacts.
Fig.~\ref{fig:noise_line} shows the evolution of the mean values of classifier-free guidance noise throughout the generation process: $s\cdot (Condition - Uncondition)$, where ``Condition" and ``Uncondition" represent the predicted noise with and without text {conditioning}, and $s$ is {the guidance scale.} We can observe that over timesteps 0 to 400, the variation in guidance noise decreases significantly, indicating that the fundamental semantics have been established. 
Based on the ablation study detailed in the Appendix Sec.~B.2, we set {$t_d \in [200, 300]$} to balance watermark robustness and image quality.

\paragraph{Watermark Decoding}
Following SSL~\cite{fernandez2022watermarking}, we employ a pre-trained image feature extractor $\mathcal{D}_{msg}$ (\eg DINO~\cite{caron2021emerging}) as our message decoder. 
{Given a watermarked image $x_0^*$, we compute its embedding $E_w=\mathcal{D}_{msg}\left(x_0^*\right) \in \mathbb{R}^{1\times D}$, and denote the secret $k$‑bit message as $m=\left(m_1,\dots,m_k \right) \in \{-1, 1\}^k$. We predefine a set of carrier vectors $\{a_i\}_{i=1}^k$, $a_i\in\mathbb R^D$, each initialized by whitening on a large natural‑image dataset to ensure that decoding on arbitrary (non‑watermarked) images yields i.i.d. Bernoulli(0.5) bits. The recovered message is then:
}
\begin{equation}
\label{message_d}
   \hat{m} = \left[ sign\left(E_{w}\cdot a_1^\top\right), \cdots, sign\left(E_{w}\cdot a_k^\top\right)\right],
\end{equation}

During training, the watermark decoding loss is defined as the hinge loss with margin $\mu \geq 0$ on the projections:
\begin{equation}
\mathcal{L}_{msg} = \frac{1}{k} \sum_{i=1}^{k} max(0,(\mu- (E_{w} \cdot a_i^\top) \cdot m_i)).
\end{equation}

\subsection{Balancing Robustness and Image Quality}
\label{Tradeoff}
\paragraph{Quality‑Preserving Components}
{To minimize watermarking’s impact on visual fidelity, we propose three complementary components: \textit{watermark initialization}, \textit{embedding strategy}, and \textit{regularization loss}. Our optimization targets two criteria: (i) the latent distribution before and after watermark embedding remains as close as possible; and (ii) the embedded watermark follows a low‑variance Gaussian profile, as diffusion models are well trained to handle small Gaussian perturbations.}


Based on the above design principles, we initialize both the structure watermark and detail watermark as $w_s^{init},w_d^{init}\sim \mathcal{N}\left( 0, 0.01 \right)$. For the structure watermark, since it is embedded into the initial diffusion latent $x_T\sim \mathcal{N}\left(0, \mathbf I\right)$, 
{we apply a two‑step normalization within the embedding operator \(\mathcal{F}_s\) to preserve unit variance:}
\begin{equation}
x_{T}^{w}= w_s + \sqrt{\frac{\text{var}\left( x_{T} \right) -\text{var}\left( w_s \right)}{\text{var}\left( x_{T} \right)}} \cdot x_{T},
\end{equation}
\begin{equation}
x_{T}^{w} = \sqrt{\frac{\text{var}\left( x_T \right)}{\text{var}\left( x_{T}^{w} \right)}} \cdot x_{T}^{w},
\end{equation}
where $\text{var}(\cdot)$ indicates the variance of data. The derivation and proof can be found in the Appendix Sec.~A. Additionally, we impose an L2 regularization to ensure that the mean of the watermarked initial diffusion latent remains {nearly unchanged} to its original value before watermarking:
\begin{equation}
  \mathcal{L}_{init}=\mathcal{L}_{mean}\left(x_T^w, x_T\right) = \left(\text{mean}\left(x_T^w\right)  - \text{mean}\left(x_T\right)\right) ^2,
\end{equation}
where $\text{mean}(\cdot)$ indicates the mean of data.


{For the detail watermark, we also aim to minimize the impact of the embedding operator \(\mathcal F_d\) on the DDIM sampling. By Eq.~\ref{eq:ddim_sampling}, the reverse process is robust to small Gaussian perturbations \(\sigma_t\epsilon_t\). Thus, at \(t = t_d\) we replace the term \(\sigma_{t_d}\epsilon_{t_d}\) with the detail watermark \(w_d\sim\mathcal N(0,0.01)\), initializing  \(\sigma_{t_d}=0.1\); for all other timesteps we use \(\sigma_t=0\).}

In addition, we further introduce losses to separately constrain the watermarks' low-order statistics (mean and variance) and high-order statistics (kurtosis and skewness), ensuring they remain statistically similar to the small initial Gaussian noise, given by:
\begin{equation}
 \begin{split}
  \mathcal{L}_{low} = \mathcal{L}_{mean}(w_s, w_s^{init}) + \mathcal{L}_{var}(w_s, w_s^{init}) \\ + \mathcal{L}_{mean}(w_d, w_d^{init}) + \mathcal{L}_{var}(w_d, w_d^{init}),
  \end{split}
\end{equation}
\vskip -0.1in
\begin{equation}
  \mathcal{L}_{high} = \mathcal{L}_{kur}(w_s) + \mathcal{L}_{kur}(w_d) + \mathcal{L}_{ske}(w_s) +  \mathcal{L}_{ske}(w_d),
\end{equation}
where $\mathcal{L}_{mean}(\cdot,\cdot)$ and $\mathcal{L}_{var}(\cdot,\cdot)$ indicates the L2 mean and variance loss. $\mathcal{L}_{kur}(x)=\left( \frac{1}{n}\sum_{i=1}^n{\left( \frac{x_i-mean\left( x \right)}{std\left( x \right)} \right) ^4-3} \right) ^2$ is the Kurtosis loss and $\mathcal{L}_{ske} = \left( \frac{1}{n}\sum_{i=1}^n{\left( \frac{x_i-mean\left( x \right)}{std\left( x \right)} \right) ^3} \right) ^2
$ is the Skewness loss. These two high-order losses constrain the shape of the watermark distribution.

\paragraph{Final Objective}
The final optimization objective is defined as a weighted combination  of the watermark decoding loss and the image‑quality constraint terms:
\begin{equation}
\mathcal{L} = \lambda_{msg}\mathcal{L}_{msg} +\lambda_{init}\mathcal{L}_{init} + \lambda_{low}\mathcal{L}_{low} + \lambda_{high}\mathcal{L}_{high},
\end{equation}
where $\lambda_{msg}$, $\lambda_{init}$, $\lambda_{low}$  and $\lambda_{high}$  are hyperparameters that balance the respective loss components.

\subsection{Optimizing with Adjoint Sensitivity Method}
{The DDIM sampler~\cite{song2020denoising} can be interpreted as an ordinary‑differential‑equation (ODE) solver.} Our objective is to minimize $\mathcal{L}$ with respect to the watermark $w$. For simplicity, we merge $w_s$ and $w_d$ into a unified notation $w$. {We optimize the watermark vector \(w\) by minimizing:}
\begin{align}
\begin{split}
    \mathcal{L}\left(w\right) &= \mathcal{L}\left(x_T+\int_T^0 f\left(x_t,t,c,w\right) dt  \right) \\
    &= \mathcal{L}\left(\text{ODESolve}\left(x_T,f,T,0,w\right)\right),
\end{split}
\end{align}
where $f$ predicts the denoising residuals, incorporating operations such as denoising noise prediction, classifier-free guidance, and scheduler scaling.
A straightforward optimization approach is to back‑propagate through the DDIM solver. However, this requires storing the entire computation graph during DDIM inference, leading to GPU memory consumption proportional to the number of inference steps, $O\left(N\right)$.
To address this, we adopt the Adjoint Sensitivity Method introduced in~\cite{chen2018neural} to compute the gradient of $\mathcal{L}$ with respect to $w$, which reduces memory cost to $O\left(1\right)$. The key idea is to compute gradients by solving a second, adjoint ODE backward in time. 
First, we define three interdependent quantities: $x_t$ is the intermediate latents at timestep $t$; $a_t=\frac{\partial \mathcal{L}}{ \partial x_t}$, is the gradient of $\mathcal{L}$ \textit{w.r.t} $x_t$; $\frac{\partial \mathcal{L}}{\partial w}$ is the gradient of $\mathcal{L}$ \textit{w.r.t.} $w$, which is also our target. The dynamics of these three quantities can be defined by the following equations:
\begin{align}
\begin{split}
\frac{dx_t}{dt}
&= f\bigl(x_t,t,c,w\bigr), \\
\frac{d a_t}{dt}
&= -a_t^\top\frac{\partial f(x_t,t,c,w)}{\partial x_t}, \\
\frac{\partial \mathcal{L}}{\partial w}
&= \int_0^T a_t^\top\frac{\partial f\left(x_t,t,c,w\right)}{\partial w}dt.
\end{split}
\label{eq:inv_dynamics}
\end{align}
Subsequently, by making a single call to the ODE solver, we simultaneously perform backward integration along the diffusion path from timestep $0$ to $T$ for all three quantities, ultimately obtaining the gradient of $\mathcal{L}$ with respect to $w$:
\begin{equation}
    [x_T, a_T, \frac{\partial \mathcal{L}}{\partial w}] =  \text{ODESolve}\left(s_0, \text{dynamics},0,T,w\right)
\end{equation}
where $s_0 = [x_0, a_0, \mathbf{0}_w]$ is the initial state of the three quantities, dynamics are 
$\left[f,-a_t^\top \frac{\partial f}{\partial x_t},- a_t^\top \frac{\partial f}{\partial w}\right]$ 
defined in Eq.~\ref{eq:inv_dynamics}.

\section{Experiments}
\label{sec:experiments}
\subsection{Experimental Setup}
\noindent \textbf{Model and Dataset.} 
{We adopt widely-used StableDiffusion‑v2.1~\cite{rombach2022high} as our generative model, and use the Stable‑Diffusion‑Prompts dataset~\cite{GustavostaStableDiffusionPromptsDatasets2023} as the source of text prompts.}

\noindent \textbf{Evaluation Metrics.} To evaluate robustness, we use bit accuracy as a metric and calculate the true positive rate (TPR) corresponding to a fixed false positive rate (FPR), which is set at $10^{-6}$, to assess the degradation of secret messages under various attacks. For image quality evaluation, we use the FID~\cite{heusel2017gans} to assess the fidelity of the watermarked image distribution and the CLIP score~\cite{radford2021learning} to measure the alignment between the generated images and their corresponding text prompts.

\noindent \textbf{Implementation Details.}
For the diffusion model, we apply the DDIM~\cite{song2020denoising} scheduler with 20 denoising steps to generate 1,000 images at $512\times512$ resolution in the main experiments.  We embed 48-bit secret messages ($k = 48$) into each image, and the pre-defined key carrier's dimension is 2048 ($D=2048$). The detail watermark is injected at step 251 ($t_{d}=251$, $15^{th}$ step). The loss weights $\lambda_{msg}$, $\lambda_{init}$, $\lambda_{low}$  and $\lambda_{high}$ are set to 0.1, 100, 1000, and 100, respectively. Inspired by SSL~\cite{fernandez2022watermarking}, DINO~\cite{caron2021emerging} is used as the pre-trained message decoder.  We employ the Adam~\cite{kingma2014adam} optimizer with 1,200 optimization rounds, and the learning rate is 0.002.

\subsection{Robustness of Watermark}

\begin{table*}[t]
\centering
\small
\setlength{\tabcolsep}{3pt}
\begin{tabular}{@{\hspace{2pt}}c X c@{\hspace{2.5pt}}c *{5}{|c@{\hspace{2.5pt}}c}}
    \toprule
   &  \multicolumn{12}{c}{\textbf{Various Attack}} \\
    \cmidrule(ll){2-13}
    &  \multicolumn{2}{c}{\textbf{None}} 
    & \multicolumn{2}{c}{\textbf{Geometric}} 
    & \multicolumn{2}{c}{\textbf{Valuemetric}} 
    & \multicolumn{2}{c}{\textbf{Editing}} 
    & \multicolumn{2}{c}{\textbf{Regeneration}} 
    & \multicolumn{2}{c}{\textbf{Average}} \\
    \textbf{Method} & Bit Acc. & TPR & Bit Acc. & TPR  & Bit Acc. & TPR  & Bit Acc. & TPR  & Bit Acc. & TPR  & Bit Acc. & TPR \\ 
    \midrule
    DwtDct~\cite{cox2007digital} & 0.828	& 0.576	& \underline{0.501}	& \underline{0.000}	& \underline{0.509}	& \underline{0.363}	& \underline{0.719}	&\underline{0.256} &  \underline{0.494}	& \underline{0.000}	&  \underline{0.573} & \underline{0.125} \\ 
    DwtDctSvd~\cite{cox2007digital} & \textbf{1.000}	& \textbf{1.000}	& \underline{0.468}	&  \underline{0.000}	& \underline{0.701}	&  \underline{0.405}	& 0.837	& 0.671	& \underline{0.605}	& \underline{0.022} & \underline{0.679} & \underline{0.340}\\ 
    RivaGAN*~\cite{zhang2019robust}    & 0.994   & 0.994  & \underline{0.742}   & \underline{0.492}   & 0.974	& 0.966	   & 0.914   & 0.775   & \underline{0.570}     & \underline{0.003} &0.835 &   0.641\\
    SSL Watermark~\cite{fernandez2022watermarking}   & \textbf{1.000}	& \textbf{1.000}	& 0.996	& 0.998 & 0.989	& 0.994 &	0.922	& 0.750 & \underline{0.596}	&\underline{0.005}  &0.906 & 0.763\\ 
    Stable Signature~\cite{fernandez2023stable}   & 0.995	& 0.998  & 0.810   & 0.496   & 0.824	& 0.724	  & \underline{0.253}   &\underline{0.498}   & \underline{0.605}    & \underline{0.011} & 0.757 &  0.509   \\
    Gaussian  Shading*~\cite{yang2024gaussian}  & \textbf{1.000}	& \textbf{1.000}	& \underline{0.634}	& \underline{0.250}	& \textbf{0.998}	& 0.997 &	0.870	& 0.750 & \textbf{0.986}	& \textbf{0.958} & 0.880 &0.756 \\ 
    AquaLoRA~\cite{feng2024aqualora}  & 0.963 & 0.979	& \underline{0.690}	& \underline{0.271}	& 0.954	& 0.973 &	0.858 	& 0.702 &  0.930	& 0.955  &  0.866 & 0.741  \\ 
    \textbf{OptMark} {\scriptsize(ours)} & \textbf{1.000}	& \textbf{1.000}	& \textbf{0.998}	& \textbf{1.000}	& \textbf{0.998}	& \textbf{1.000}	& \textbf{0.990}	& \textbf{0.979}	& 0.923	& 0.872 & \textbf{0.983} & \textbf{0.972} \\ 
    \bottomrule
\end{tabular}
\vskip -0.1in
\caption{Performance of multi-bit different watermarking methods under various attacks on DiffusionDB~\cite{GustavostaStableDiffusionPromptsDatasets2023}. ``Average'' indicates calculating the average score across cases under sixteen different attacks and the no-attack (``None''). ``*'' indicates that Gaussian Shading~\cite{yang2024gaussian} and RivaGAN~\cite{zhang2019robust} embed 64-bit and 32-bit hidden messages respectively, whereas all other methods are compared under the condition of embedding 48-bit messages. The \underline{underline} indicates poor robust performance with Bit Acc. $< 0.75$ and TPR $< 0.5$.}
\label{table:comparison}
\vskip -0.1in
\end{table*}
 The various attack methods that we implement can be divided into four categories: \textit{geometric attack} (horizontal flip, random rotation of 40 degrees, resizing of 60\%, and center cropping of 60\%), \textit{valuemetric attack} (color jitter with brightness 0.5, Gaussian blur with radius 11, contrast adjustment to 0.5, 50\% JPEG compression, and saturation adjustment to 1.5),  \textit{editing attack} (Meme format, random erase with area ratio of 0.1, text overlay, and InstructPix2Pix~\cite{brooks2023instructpix2pix}) and \textit{regeneration attack} (two types of VAE regeneration attacks~\cite{balle2018variational, cheng2020learned} from the CompressAI library~\cite{begaint2020compressai} with a compression factor of 3, and a diffusion regeneration attack performed with 60 denoising steps~\cite{zhao2023invisible}.) The processed samples after diverse attacks are shown in the Appendix Sec.~B.5.

\subsubsection{Multi-bit Methods Comparison}
{For multi‑bit watermarking, we evaluate our OptMark against seven baselines: DwtDct~\cite{cox2007digital}, DwtDctSvd~\cite{cox2007digital},  RivaGAN~\cite{zhang2019robust}, SSL Watermark~\cite{fernandez2022watermarking}, Stable Signature~\cite{fernandez2023stable}, Gaussian  Shading~\cite{yang2024gaussian}, and AquaLoRA~\cite{feng2024aqualora}. {Except for Gaussian Shading and AquaLoRA}, which embed watermarks in the diffusion latent space, all other methods operate in pixel space.}
Tab.~\ref{table:comparison} shows the watermark robustness comparison between other methods and our OptMark. We find that SSL Watermark~\cite{fernandez2022watermarking} exhibits strong robustness against attacks, except for generative attacks, making it stand out among all pixel-space embedding methods. However, it is worth noting that all pixel space embedding methods exhibit little to no resistance to generative attacks. In contrast, the diffusion space embedding method Gaussian Shading~\cite{yang2024gaussian} and {AquaLoRA~\cite{feng2024aqualora}} exhibits strong robustness against regeneration attacks but is rendered ineffective when facing geometric attacks. Unlike them, our OptMark is a highly comprehensive approach that demonstrates exceptional robustness against various attacks without evident weaknesses, achieving SOTA performance. A more detailed experiment on the performance of various methods against different attacks can be found in the Appendix Sec.~B.6.

\subsubsection{Zero-bit Methods Comparison}
{For zero-bit watermarking, we compare our OptMark with Tree-Rings~\cite{wen2023tree}, RingID~\cite{ci2024ringid}, and WIND~\cite{arabi2024hidden}.  All of these approaches embed semantic‑level watermarks within the diffusion latent space.  Consistent with the standard evaluation for zero‑bit schemes, we report all results under TPR@FPR=$1\%$, with results shown in Tab.~\ref{zero_qua_tab}. Compared to alternative methods, our approach demonstrates superior robustness against all attack types, exhibiting no vulnerability to any specific attack and achieving the best overall robustness performance.
}
\begin{table}[t!]
\centering
\small
\setlength{\tabcolsep}{5pt}
{\begin{tabular}{ccccccc}
\hline
\textbf{Method}  & \textbf{{None}}  & \textbf{{Geo.}}  & \textbf{{Valu.}} & \textbf{{Edit.}}  & \textbf{{Regen.}}& \textbf{Avg.}\\ \hline
Tree-Ring &\textbf{1.000} & 0.773 & 0.970 &0.765 &0.953 &0.874  \\
RingID &\textbf{1.000} &0.750 & 0.999& 0.717& 0.814&  0.841 \\
WIND &\textbf{1.000} &0.985 & 0.976 & 0.748 & \textbf{1.000} & 0.930  \\
{\textbf{OptMark}} &\textbf{1.000} &\textbf{1.000} & \textbf{1.000} & \textbf{1.000}&  0.993 & \textbf{0.999} \\     \hline
\end{tabular}}
\vskip -0.1in
\caption{Performance of zero-bit different watermarking methods under various attacks.
}
\label{zero_qua_tab}
\vskip -0.1in
\end{table}

\subsection{Quality of Watermarked Image} 

\begin{figure*} [t!]
	\centering
	\includegraphics[width=\linewidth]{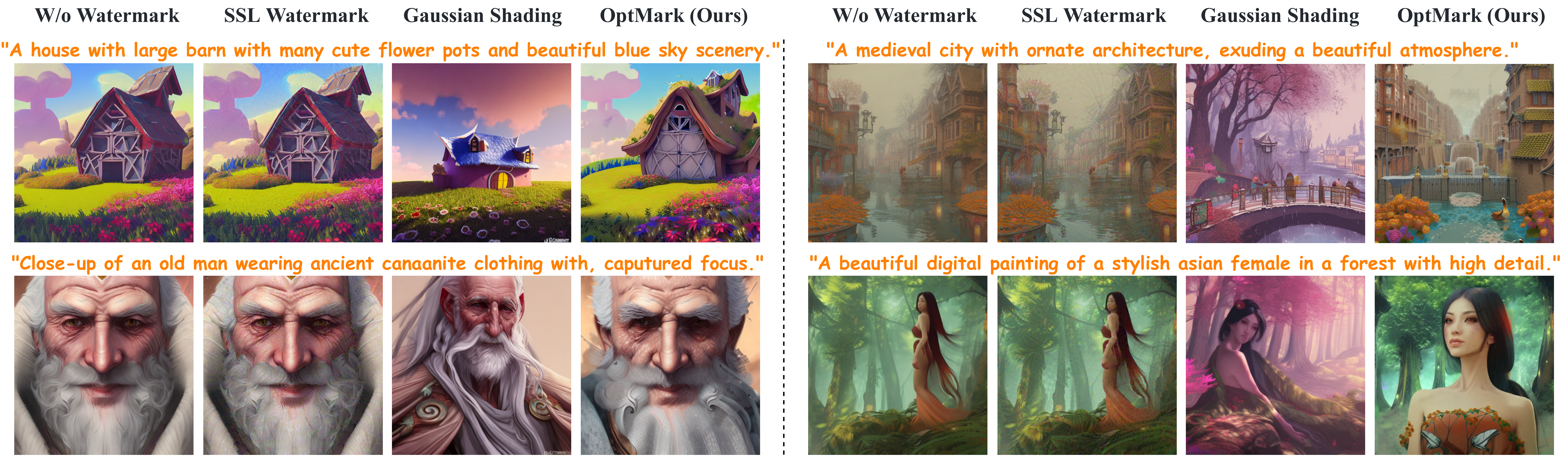}
\vskip -0.1in
	\caption{Qualitative comparison of image quality between SSL Watermark~\cite{fernandez2022watermarking}, Gaussian Shading~\cite{yang2024gaussian}, and our proposed OptMark.}
	\label{fig:result}
\vskip -0.1in
\end{figure*}

\begin{table}[t!]
\centering
\small
\setlength{\tabcolsep}{2pt}

{\begin{tabular}{ccc}
\hline
\textbf{Method}  & {$\textbf{FID}\downarrow$}  & {$
\textbf{CLIP Score}\uparrow$} \\ \hline
    w/o watermark & 124.309 &  0.3686\\ \hline 
    SSL Watermark~\cite{fernandez2022watermarking}  & 128.053 & 0.3555\\
    Gaussian  Shading~\cite{yang2024gaussian} & 127.756 & \textbf{0.3646} \\
    \textbf{OptMark} {\scriptsize(ours)}	  & \textbf{127.378} & 0.3630 \\ \hline
\end{tabular}}
\vskip -0.1in
\caption{Quantitative analysis of the watermarked image quality. ``w/o watermark'' indicates the baseline using images generated by Stable Diffusion~\cite{rombach2022high} without watermarks.
} 
\label{qua_tab}
\vskip -0.1in
\end{table}

The qualitative image quality comparison is shown in Fig.~\ref{fig:result}. SSL Watermark~\cite{fernandez2022watermarking} introduces noticeable artifacts due to the disturbance added in the pixel space. In contrast, Gaussian Shading~\cite{yang2024gaussian} only adds the watermark to the initial latent in the diffusion model without impacting the denoising process, resulting in image quality comparable to that of images without watermark. Although our OptMark injects two watermarks (structure and detail watermark) during the denoising process, the image quality remains unaffected compared to Gaussian Shading and images without watermark, and the semantic representation stays consistent with the corresponding text prompt, demonstrating the effectiveness of our method.

For a quantitative comparison of image quality, we compare the FID~\cite{heusel2017gans} and CLIP Score~\cite{radford2021learning}. The FID is evaluated on the MS-COCO-2017 dataset~\cite{lin2014microsoft}. As shown in Table~\ref{qua_tab}, our OptMark achieves the best performance in FID, indicating the closest alignment to the real data distribution. Furthermore, it demonstrates a CLIP Score comparable to Gaussian Shading~\cite{yang2024gaussian}.

\subsection{Ablation Study}
In this section, to more clearly illustrate the changes in quality metrics, we introduce  $\Delta_\textbf{FID}$ and $\Delta_\textbf{CLIP-Score}$, both of which are relative values compared to the baseline, \textit{i.e.}, ``w/o watermark''. All training iterations in the following ablation studies are set to 1,200.

\begin{figure} [t!]
	\centering
	\includegraphics[width=0.95\linewidth]{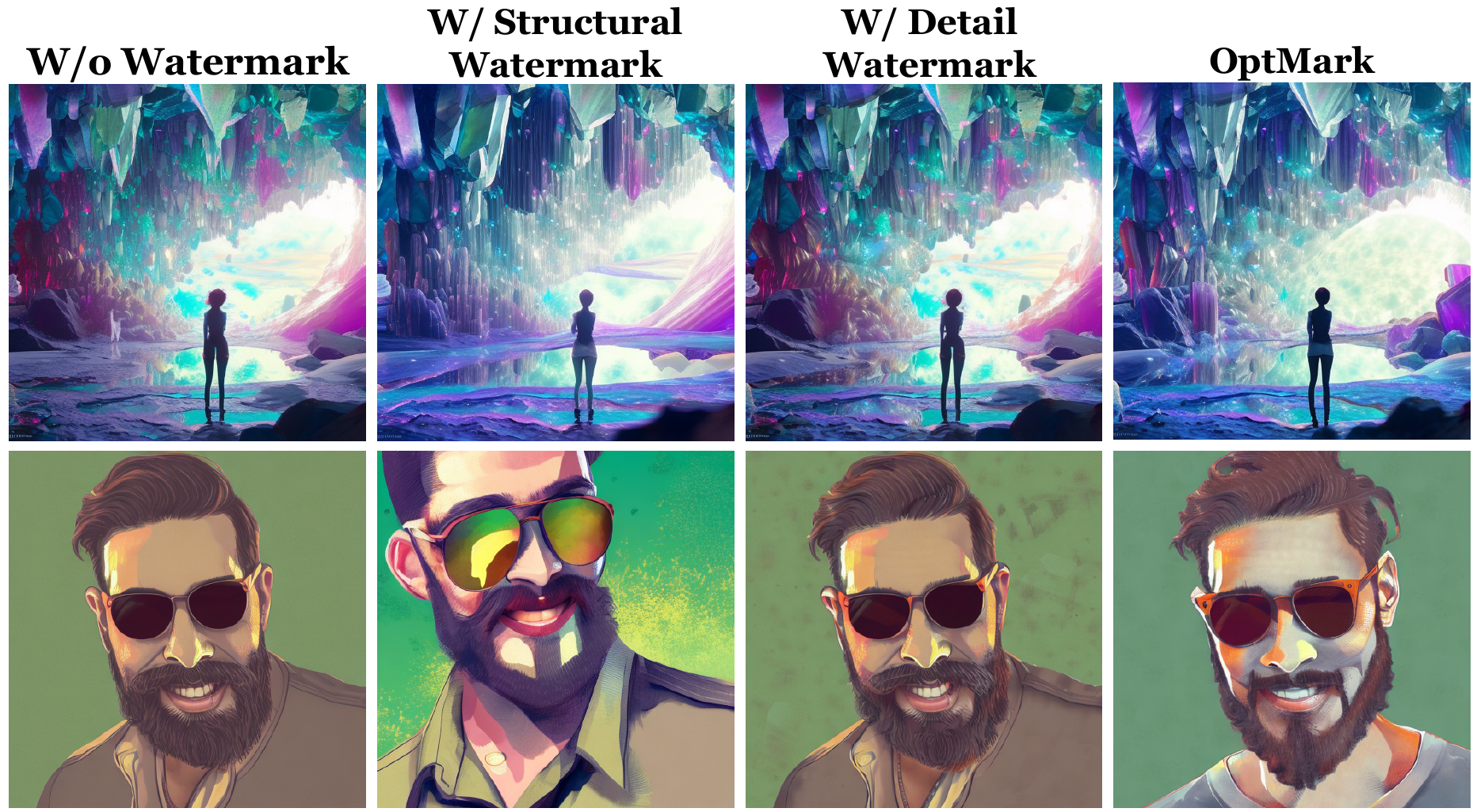}
    \vskip -0.1in
	\caption{Visualization of the generated images adding different watermarks.}
      \vskip -0.05in
	\label{fig:watermark_effect}  
\end{figure}

\noindent \textbf{Effect of Dual Watermarks} 
We conduct both quantitative and qualitative analyses to demonstrate the necessity of combining the structure watermark and the detail watermark. The quantitative results are shown in Tab.~\ref{tab:watermark_effect}. From the table, it can be observed that the structure watermark, introduced during the structure formation stage, demonstrates stronger robustness against regeneration attacks compared to the detail watermark, which is introduced in the detail formulation stage.
However, the structure watermark's convergence is relatively slower, and optimization over 1200 iterations is insufficient for full convergence. As a result, its performance against conventional attacks is weaker than that of the detail watermark. The combination of both can accelerate convergence and result in a more robust performance under various attacks. For qualitative analysis, as shown in Fig.~\ref{fig:watermark_effect}, the introduction of the detail watermark closer to the final image generation stage makes it prone to issues similar to those encountered in pixel-level watermarking methods (e.g., SSL~\cite{fernandez2022watermarking}), such as the appearance of artifacts. In contrast, the structure watermark does not exhibit this problem. Since it is introduced at the semantic level, it also leads to some visual differences compared to the original image without watermarks. Furthermore, the combination of both watermarks helps mitigate artifacts.

\begin{table}[t!]
\centering
\setlength{\tabcolsep}{5pt}
\small
\begin{tabular}{cccccc}
\toprule
 & & 
\multicolumn{2}{c}{\textbf{Other Attacks}} & \multicolumn{2}{c}{\hspace{-4pt}\textbf{Regeneration Attack}}   \\
\multirow{-2}{*}{\textbf{Structure}}& \multirow{-2}{*}{\textbf{Detail}} & {Bit Acc.} & TPR  & {Bit Acc.} & {TPR} \\ \midrule
$\usym{2713}$ & $\usym{2717}$   & 0.961 &0.935 &  0.834 & 0.567 \\
$\usym{2717}$  & $\usym{2713}$  &0.984 & 0.990 & 0.794 & 0.407  \\ 
 $\usym{2713}$  &  $\usym{2713}$  &\textbf{0.993}&  \textbf{1.000}& \textbf{0.923}& \textbf{0.872} \\ \bottomrule
\end{tabular}
\vskip -0.1in
\caption{Effect of different watermarks. ``Structure" and ``Detail" refer to the structure watermark and detail watermark, respectively. ``Other Attacks'' encompasses various attacks, including geometric, valuemetric, and editing attacks.}
\vskip -0.1in
\label{tab:watermark_effect}
\end{table}

\begin{figure} [t!]
	\centering
	\includegraphics[width=\linewidth]{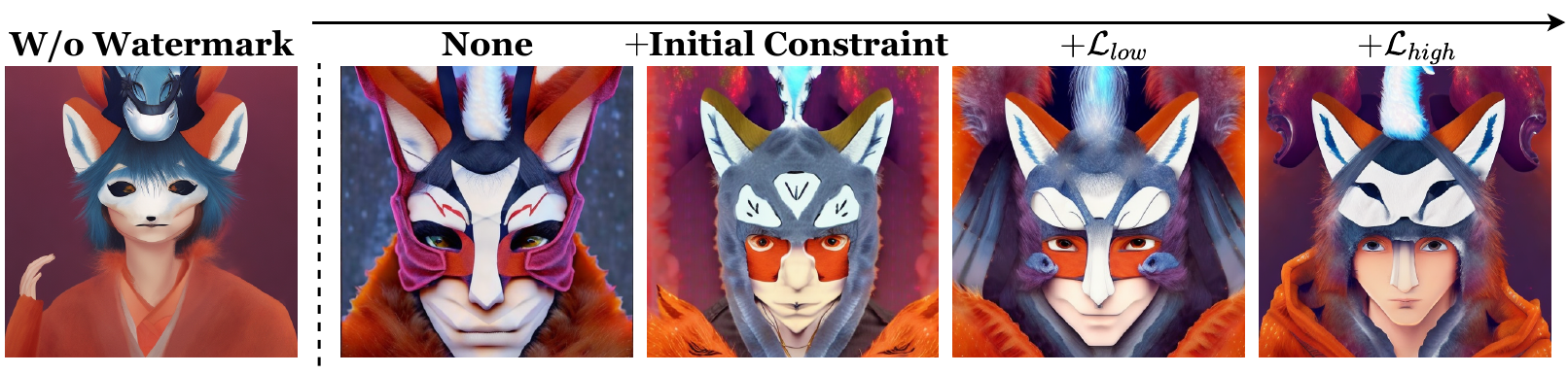}
\vskip -0.1in
	\caption{Visualization of our quality-driven constraint methods applied to the watermarked images. ``Initial Constraint'' includes the normalization step in $\mathcal{F}_s$ and $\mathcal{F}_d$, and the loss $\mathcal{L}_{init}$.}
	\label{fig:quality_abl}
    \vskip -0.05in
\end{figure}
\noindent \textbf{Effect of Image Quality Constraints.} 
To assess the effectiveness of the proposed image quality constraints, we perform an ablation study on each individual component.  The results are presented in Fig.~\ref{fig:quality_abl} and Tab.~\ref{tab:quality_abl}. Note that ``Initial Constraint'' includes the normalization step in $\mathcal{F}_s$ and $\mathcal{F}_d$, and the loss $\mathcal{L}_{init}$. From a qualitative perspective, as shown in Fig.~\ref{fig:quality_abl}, the realism and quality of the generated images progressively improve with the introduction of each quality-driven constraint. From a quantitative perspective, as shown in Tab.~\ref{tab:quality_abl}, our constraints achieve a significant improvement in FID with only a minimal loss in robustness.

\noindent \textbf{Effect of Adjoint Method.} 
In experiments, we find that under the DDIM setting with 20 inference steps, the GPU memory consumption of the naive optimization is about 52 GB, while for 30 steps, it increases to around 76 GB. Using the adjoint method, the memory consumption can be uniformly reduced to just 9 GB, making it feasible to scale to larger inference steps and more complex diffusion models.

\begin{table}[t!]
\centering
\small
\setlength{\tabcolsep}{2pt}
\begin{tabular}{ccccccc}
\toprule
  & \multicolumn{2}{c}{\textbf{Robustness}} & \multicolumn{2}{c}{\textbf{Image Quality}}   \\
 \multirow{-2}{*}{\textbf{Method}} & {Bit Acc.} & TPR  & $\Delta_\textbf{FID} \downarrow$ & $\Delta_\textbf{CLIP-Score}\uparrow$\\ \midrule
 None  &\textbf{0.985}  &\textbf{0.975} &  {9.779} & -0.0085 \\
  Init. Cons.  &0.984 & 0.974 & 7.928 & -0.0083  \\  
 Init. Cons. + $\mathcal{L}_{low}$  &0.984 & 0.973 & 4.434 & -0.0059  \\  
   Init. Cons. + $\mathcal{L}_{low}$+ $\mathcal{L}_{high}$  &{0.983}&  {0.972}&  \textbf{3.069}&  \textbf{-0.0056}\\ \bottomrule
\end{tabular}
\vskip -0.1in
\caption{Effect of watermarks' different initialization. The robustness results here refer to the average scores calculated under four types of attacks and the non-attack scenario. Note that “Init. Cons.” denotes the Initial Constraint.}
\vskip -0.1in
\label{tab:quality_abl}
\end{table}

\section{Conclusion}

This paper presents OptMark, a robust watermarking framework based on inference-time optimization. We propose a dual-watermark mechanism to enhance robustness, design a tailored objective and regularization scheme to preserve image fidelity, and integrate the adjoint sensitivity method for constant‑memory gradient computation. Extensive experiments show that OptMark delivers SOTA robustness across a diverse range of common attacks.

\bibliography{aaai2026}
\clearpage

\twocolumn[
    \centering
    \textbf{\LARGE OptMark: Robust Multi-bit Diffusion Watermarking via \\ Inference Time Optimization}\\
    \vspace{1.2em}
    \Large Supplementary Material
    \vspace{1.2em}
    ]

\appendix


\maketitle

In this Appendix, we provide additional content organized as follows:
\begin{itemize}
  \item \textbf{Sec.~\ref{Derivation}} discusses the derivation of the variance constraint at step $T$.
  \item \textbf{Sec.~\ref{Experimental Results}} provides more experimental results, including:
  \begin{itemize}
    \item Sec.~\ref{inital} Initialization of watermarks.
    \item Sec.~\ref{placement detailed} Discussion on optimal placement of detailed watermarks.
    \item Sec.~\ref{different decoder} Results under different watermark decoders.
    \item Sec.~\ref{infer steps} Results under different inference steps.
    \item Sec.~\ref{various attaks} Sample outputs under various attacks.
    \item Sec.~\ref{Quantitative Results} Detailed quantitative results.
    \item Sec.~\ref{Forgery and Removal Attacks} Robustness to forgery and removal attacks.
    \item Sec.~\ref{Empty Prompt} Empty prompt example.
    \item Sec.~\ref{Training Steps on Robustness} The impact of training steps on robustness.
    \item Sec.~\ref{Generality} Generality of OptMark across different diffusion samplers.
    \item Sec.~\ref{Qualitative Results} Additional qualitative examples.
  \end{itemize}
  \item  \textbf{Sec.~\ref{Limit}} discusses the limitations of our current approach.
\end{itemize}

\section{Analytical Derivation of the Variance Constraint at Step $T$}
\label{Derivation}
The derivation of operation $\mathcal{F}_s$ for controlling the variance of output $x_{T}^{w}$, which is the combination of initial noise $x_T$ and structure watermark $w_s$, to be $\text{var}(x_T^w) \approx 1$ in Sec. 3.4 of the main paper is as follows. Given that $x_T\sim \mathcal{N}\left( 0, \mathbf{I} \right)$ and initial $w^{init}_s\sim \mathcal{N}\left( 0,{0.01} \right)$, the combination of them $x_{T}^{w}$ can be defined as:
\begin{equation}
x_{T}^{w}=   w_s + \gamma \cdot x_{T} 
\end{equation}
where $\gamma \in (0, +\infty )$ is a variable control coefficient that ensures the variance of $x_{T}^{w}$ remains constant at $\text{var}(x_T^w) \approx 1$. The reason we apply $m$ to $x_T$ instead of $w_s$ is that the variance of $w_s$ is very small. Scaling it does not effectively control the variance and would severely impact the robustness of the watermark. Our goal is to solve for $\gamma$, which has an exact solution. The derivation is as follows:
\begin{equation}
\begin{aligned}
\text{var}(x_{T}^{w}) &= \text{var}(w_s) + \text{var}(\gamma \cdot x_T) + 2 \, \text{cov}(w_s, \gamma \cdot x_T) \\
&= \text{var}(w_s) + \gamma^2 \, \text{var}(x_T) + 2 \, \text{cov}(w_s, \gamma \cdot x_T)
\end{aligned}
\end{equation}
where $\text{var}(\cdot)$ and $\text{cov}(\cdot,\cdot)$ indicates the variance and covariance calculations, respectively. To ensure that  $\text{var}(x_{T}^{w})$ equals $\text{var}(x_{T})$, the linear equation can be written as follows:
\begin{equation}
    \gamma^2 \, \text{var}(x_T) + 2 \, \text{cov}(w_s, \gamma \cdot x_T) + \text{var}(w_s) - \text{var}(x_{T}) =0
\end{equation}
To solve this linear equation, we can obtain the exact solution for $\gamma$, as follows:
\begin{equation}
\label{eq:m_sovle}
\begin{aligned}
\gamma = &\frac{\text{cov}\left ( w_s,x_T  \right )}{\text{var}\left ( x_T \right )} \pm  \\
 & \frac{\sqrt{\left ( \text{cov}\left ( w_s,x_T  \right ) \right )^2-\text{var}\left ( x_T \right )\cdot\left ( \text{var}\left ( w_s \right )- \text{var}(x_{T})\right )   } }{\text{var}\left ( x_T \right )} 
\end{aligned}
\end{equation}
The condition for real roots to exist is $\sqrt{\left ( \text{cov}\left ( w_s,x_T  \right ) \right )^2-\text{var}\left ( x_T \right )\cdot\left ( \text{var}\left ( w_s \right )- \text{var}(x_{T})\right )} > 0$,  but due to the small covariance $\text{cov}\left ( w_s,x_T  \right )$, it is difficult for real roots to occur.
Considering this situation, we assume that $w_s$ and $x_T$ are independent and identically distributed, such that their covariance $\text{cov}(w_s, x_T)$ is zero. Eq.~\ref{eq:m_sovle} can be simplified to:
\begin{equation}
    \gamma = \pm \sqrt{\frac{\text{var}\left( x_{T} \right) -\text{var}\left( w_s \right)}{\text{var}\left( x_{T} \right)}}
\end{equation}
Considering our scenario ($\gamma>0$), we only take the positive root, which leads to Eq. 5 in the main paper. The combination $x_T^w$ can be represented as :
\begin{equation}
\label{eq6}
x_{T}^{w}= w_s + \sqrt{\frac{\text{var}\left( x_{T} \right) -\text{var}\left( w_s \right)}{\text{var}\left( x_{T} \right)}} \cdot x_{T}
\end{equation}
However, we observe that in the later stages of training, $w_s$ and $x_T$
does not follow the assumption of independent and identically distributed variables. As a result, Eq.~\ref{eq6} alone is insufficient to constrain the variance of $x_T^w$ to $\text{var}(x_T) \approx 1$. To address this, we introduce an additional scaling step to ensure that $\text{var}(x_T^w)$ matches $\text{var}(x_T)$, as follows:
\begin{equation}
x_{T}^{w} = \sqrt{\frac{\text{var}\left( x_T \right)}{\text{var}\left( x_{T}^{w} \right)}} \cdot x_{T}^{w} 
\end{equation}
We don't initially scale the direct combination $x_T'^w =w_s + x_T$ as a whole because we don't want to apply a large-scale transformation to $w_s$, which can affect the robustness of the watermark. Meanwhile, since adding a watermark is an entropy-increasing process, appropriately compressing $x_T$ (usually $\gamma<1$) to make room for $w_s$ can help with the learning of the watermark. 

\section{More Experimental Results}
\label{Experimental Results}
\subsection{Initialization of Watermarks} 
\label{inital}
\begin{figure} [t!]
	\centering
	\includegraphics[width=0.99\linewidth]{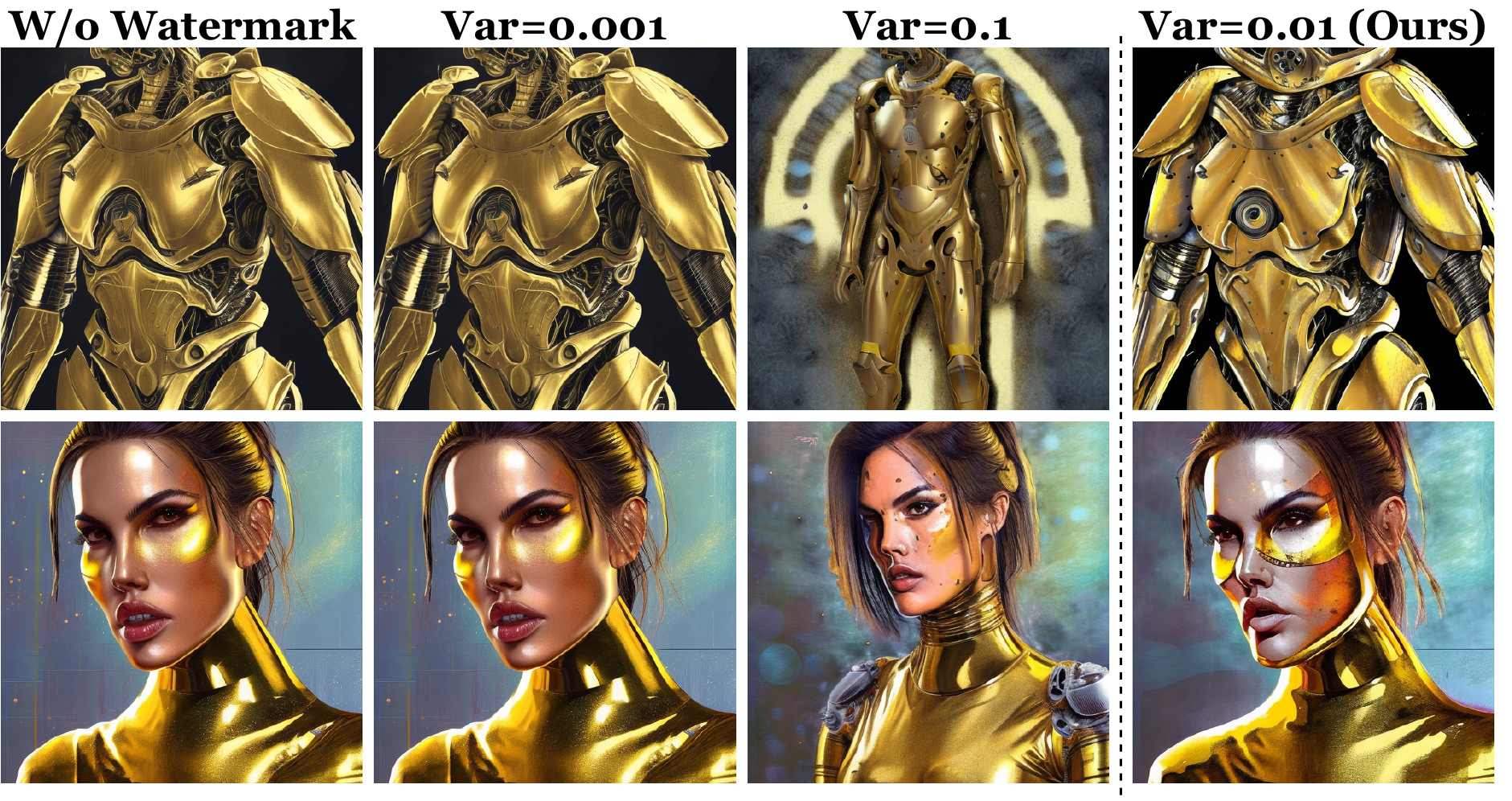}
\vskip -0.05in
	\caption{Visualization of the watermarked images with different initial watermark variance. ``Var'' indicates the initial variance of watermarks.}
	\label{fig:var_init}
\vskip -0.05in
\end{figure}

The watermarks are initialized following a Gaussian distribution, with a mean of 0 and various possible choices for the variance. Intuitively, a smaller initial variance results in a lesser impact on the generated image. As shown in Tab.~\ref{tab:watermark_init}, an excessively small initial variance, such as when $variance = 0.001$, makes it difficult for the watermark to be optimized in any case, leading to a final outcome with almost no robustness, where the generated result is nearly identical to the baseline. A larger initial variance helps improve the robustness of the optimized images but also reduces the quality of the generated images. To find a balance, we set the variance to 0.01. Qualitative comparisons are shown in Fig.~\ref{fig:var_init}.

\begin{table}[t!]
\centering
\small
\setlength{\tabcolsep}{1mm}

\begin{tabular}{ccccccc}
\toprule
  & \multicolumn{2}{c}{\textbf{Robustness}} & \multicolumn{2}{c}{\textbf{Image Quality}}   \\
 \multirow{-2}{*}{\textbf{Method}} & {Bit Acc.} & TPR  & $\Delta_\textbf{FID} \downarrow$ & $\Delta_\textbf{CLIP-Score} \uparrow$ \\ \midrule
$variance = 0.001$   & 0.502 &0.000 &  \textbf{0.000} & \textbf{0.0000} \\
 $variance = 0.1$  &\textbf{0.985} & \textbf{0.976} & 4.620 & -0.0060  \\  \midrule
  $variance = 0.01$ (ours)  &{0.983}&  {0.972}&  3.069&  -0.0056\\ \bottomrule
\end{tabular}
\vskip -0.05in
\caption{Effect of watermarks' different initialization. The robustness results here refer to the average scores calculated under four types of attacks and the non-attack scenario.}
\vskip -0.05in
\label{tab:watermark_init}
\end{table}

\subsection{Discussion on Optimal Placement of Detailed Watermarks}
\label{placement detailed}
In Sec. 3.3 of the main paper, we discuss the injection positions of the detail watermark. Here, we further validate our reasoning through experiments conducted on 100 randomly sampled cases. We test four different time injection points under the setting of a total of 20 inference steps: $t_d=51\ (0 < t_d < 100)$, $ t_d=151\ (100 < t_d < 200)$, $t_d=251 \ (200 < t_d < 300)$, and $t_d=351 \ (300 < t_d < 400)$. 
The quantitative analysis is presented in Tab.~\ref{tab:texture_steps}. In terms of robustness performance, smaller values of $t_d$ lead to poorer performance under regeneration attacks but better performance under other attacks (geometric, valuemetric, and editing attacks) and vice versa. We believe this phenomenon occurs because smaller values of $t_d$  bring the watermark closer to the pixel level. While pixel-level watermarks exhibit weak robustness against regeneration attacks, they still maintain decent robustness against other attacks (\textit{e.g.}, SSL~\cite{fernandez2022watermarking}). We find $t_d=251$ strikes a good balance between robustness and image quality. So we choose it as the default embedding timestep for detail watermark.



\begin{table}[t!]
\centering
\small
\setlength{\tabcolsep}{5pt}
\begin{tabular}{ccccccc}
\toprule
  & \multicolumn{2}{c}{\textbf{Robustness}} & \multicolumn{2}{c}{\textbf{Image Quality}}   \\
 \multirow{-2}{*}{\textbf{Method}} & {Others} & {Regeneration}  & $\text{FID} \downarrow$ & $\text{CLIP Score} \uparrow$ \\ \midrule
$t_d= 51$   & \textbf{0.998} &  0.893 &  {251.540
} & {0.3656} \\
 $t_d= 151$  &0.997 & 0.919 &  250.979 & {0.3696} \\ 
  $t_d= 251$   &0.996 &  0.931&  \textbf{248.978} &  \textbf{0.3708}\\ 
  $t_d= 351$   &0.994 &  \textbf{0.940}&  251.272 & 0.3632\\ \bottomrule
\end{tabular}
\vskip -0.05in
\caption{Effect of choosing different positions of detail watermarks. ``Others'' refers to the average bit accuracy of our OptMark under various attacks, including geometric, valuemetric, and editing attacks. ``Regeneration'' indicates the bit accuracy of our method under regeneration attacks.}
\vskip -0.05in
\label{tab:texture_steps}
\end{table}

\subsection{Results under Different Watermark Decoder}
\label{different decoder}
To assess the influence of the watermark decoder, we evaluate DINO V1-RN50~\cite{caron2021emerging} and DINO V2-ViT-S~\cite{oquab2023dinov2}. Both are trained for 1200 steps. As shown in Tab.~\ref{tab:dino}, DINO V1 demonstrates better robustness and image quality. Therefore, we chose DINO V1-RN50 as the default watermark decoder.
\begin{table}[t!]
\centering
\setlength{\tabcolsep}{5pt}
\small
\begin{tabular}{ccccccc}
\toprule
  & \multicolumn{2}{c}{\textbf{Robustness}} & \multicolumn{2}{c}{\textbf{Image Quality}}   \\
 \multirow{-2}{*}{\textbf{Method}} & {Bit Acc.} & TPR  & $\text{FID} \downarrow$ & $\text{CLIP Score} \uparrow$ \\ \midrule
DINO V1   & \textbf{0.984} & \textbf{0.974} &  \textbf{250.782} & \textbf{0.3712} \\
  DINO V2   &0.833 & 0.575 &  252.789 &  0.3634\\ \bottomrule
\end{tabular}
\vskip -0.05in
\caption{Effect of different watermarks' detector under $1200$ training steps. The robustness results here refer to the average scores calculated under four types of attacks and the non-attack scenario.}
\vskip -0.05in
\label{tab:dino}
\end{table}

\subsection{Results under Different Inference Steps}
\label{infer steps}
\begin{figure*} [t!]
	\centering
	\includegraphics[width=\linewidth]{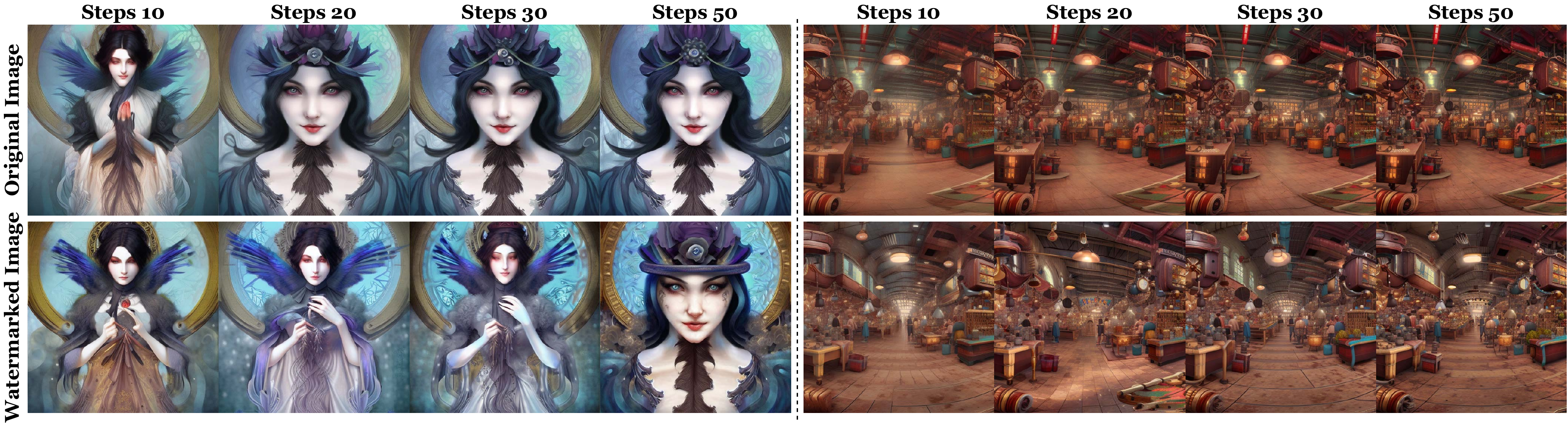}
\vskip -0.05in
\caption{Qualitative comparison under different inference steps. }
\vskip -0.05in
	\label{fig:diff_steps}
\end{figure*}

To evaluate the impact of different inference steps $T$ on our OptMark we randomly sample 100 cases for both quantitative and qualitative experiments. The quantitative results are shown in Tab.~\ref{tab:inf_steps}. From the perspective of watermark robustness, our OptMark exhibits stable performance without significant fluctuations across different inference steps.  In terms of image quality,  $T=30$ performs the best, followed by $T=20$. The qualitative results are illustrated in Fig.~\ref{fig:diff_steps}. From the figure, it can be observed that when $T=10$, the original images and watermarked images are slightly blurred and lack some high-frequency details. However, when $T\ge20$, the generated details become more stable, and the images maintain stronger integrity. Although we apply the adjoint method, the memory usage of OptMark remains consistent across different inference step sizes. However, as $T$ increases, the training time becomes longer. To achieve a balance between efficiency, image quality, and watermark robustness, we finally set $T=20$.

\begin{table}[t!]
\centering
\small
\setlength{\tabcolsep}{5pt}

\begin{tabular}{ccccccc}
\toprule
  & \multicolumn{2}{c}{\textbf{Robustness}} & \multicolumn{2}{c}{\textbf{Image Quality}}   \\
 \multirow{-2}{*}{\textbf{Method}} & {Bit Acc.} & TPR  & $\text{FID} \downarrow$ & $\text{CLIP Score} \uparrow$ \\ \midrule
$T= 10$   & 0.983 & 0.972 &  {250.088} & {0.3611} \\
 $T = 20$  &\textbf{0.985} & \textbf{0.973} & 250.830 & {0.3704} \\ 
  $T = 30$   &{0.984}&  {0.972}&  \textbf{243.949} &  \textbf{0.3713}\\ 
  $T = 50$   &\textbf{0.985}&  \textbf{0.973}&  256.621 &  0.3628\\ \bottomrule
\end{tabular}
\vskip -0.05in
\caption{Effect of watermarks' inference steps $T$. The robustness results here refer to the average scores calculated under four types of attacks and the non-attack scenario.}
\vskip -0.05in
\label{tab:inf_steps}
\end{table}

\begin{figure*} [ht!]
	\centering
	\includegraphics[width=\linewidth]{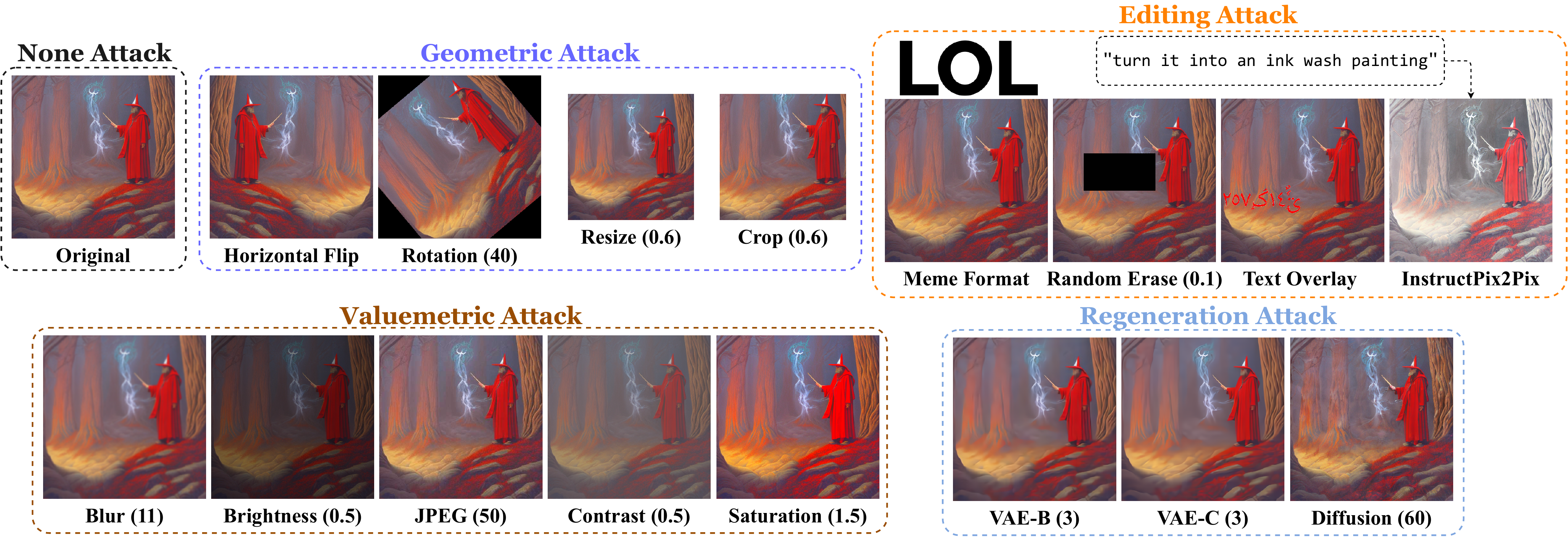}
\vskip -0.05in
\caption{Samples under different attacks.}
	\label{fig:sample_attacks}
    \vskip -0.05in
\end{figure*}

\subsection{Samples under Different Attacks}
\label{various attaks}
The visualization of samples under different attacks can be seen in Fig.~\ref{fig:sample_attacks}. Specifically, the various attack methods we implement can be categorized into four types: \textit{geometric attack} (horizontal flip, random rotation of 40 degrees, resizing of 60\%, and center cropping of 60\%), \textit{valuemetric attack} (color jitter with brightness 0.5, Gaussian blur with radius 11, contrast adjustment to 0.5, 50\% JPEG compression, and saturation adjustment to 1.5),  \textit{editing attack} (Meme format, random erase with a probability of 0.1, text overlay, and InstructPix2Pix~\cite{brooks2023instructpix2pix} with the prompt: ``turn it into an ink wash painting") and \textit{regeneration attack} (two types of VAE regeneration attacks~\cite{balle2018variational, cheng2020learned} from the CompressAI library~\cite{begaint2020compressai} with a compression factor of 3, and a diffusion regeneration attack performed with 60 denoising steps~\cite{zhao2023invisible}.) 

\subsection{Detailed Quantitative Results}
\label{Quantitative Results}
Tab.~\ref{table:detal_compare} presents the detailed detection results for all attacks. We observe that our OptMark is capable of handling all of them. In contrast, other methods exhibit at least one weakness in dealing with these attacks. For instance, nearly all pixel-level methods, including DwtDct~\cite{cox2007digital}, DwtDctSvd~\cite{cox2007digital}, RivaGAN*\cite{zhang2019robust}, SSL\cite{fernandez2022watermarking}, and Stable Signature~\cite{fernandez2023stable}, struggle with Generative attacks. Meanwhile, the semantic-level method Gaussian Shading~\cite{yang2024gaussian} and AquaLoRA~\cite{feng2024aqualora} perform poorly under Geometric attacks.

\subsection{Robustness to Forgery and Removal Attacks}
\label{Forgery and Removal Attacks}
{We assess the impact of several attacks—previously shown to be highly disruptive to other watermarking methods—on our proposed OptMark framework, including the Imprint-Forgery and Imprint-Removal Attacks from ~\cite{muller2025black}, as well as the Averaging Attack from ~\cite{yang2024steganalysis}. We evaluate robustness to all attacks on a set of 1,000 images, with results shown in Tab.~\ref{black_box attack}. The results demonstrate that the Imprint-Forgery Attack fails to forge our OptMark even after running 150 steps, achieving only {0.563} average multi-bit accuracy. In terms of Imprint-Removal Attack at 50, 100, 150 steps, our method achieves bit accuracy of {0.937}, {0.832}, and {0.791}, significantly outperforms Gaussian Shading, whose bit accuracy is consistently lower than {0.2}. Under the Averaging Attack, which averages 1,000 watermarked images, OptMark maintains a bit accuracy of 0.996, demonstrating strong robustness. 
We attribute this resilience to our per-image optimization strategy: Averaging attacks are effective only when the watermark pattern is independent of image content, whereas OptMark generates image-specific watermarks, rendering such attacks ineffective.
}
\begin{table}[t]
\resizebox{0.47\textwidth}{!}{
\begin{tabular}{llcccc}
\toprule
 &  &\multicolumn{1}{c}{G. Shad.} & \multicolumn{1}{c}{Tree-Ring} &  \multicolumn{2}{c}{\textbf{OptMark}}  \\
 Attack & Step    & Bit Acc. & TPR & Bit Acc. & TPR \\
\midrule

 \multirow[c]{3}{*}{Imprint-F} & 50 & 0.967 & 1.000 & \textbf{0.552} &\textbf{0.000} \\
 & 100 & 0.978 & 1.000  & \textbf{0.557}  & \textbf{0.000}\\
 & 150 & 0.989 &  1.000  & \textbf{0.563} & \textbf{0.000}\\ \midrule 
  \multirow[c]{3}{*}{Imprint-R} & 50 & 0.183 & 0.164 & \textbf{0.937} & \textbf{0.998} \\
 & 100 & 0.084 & 0.052  & \textbf{0.832}  & \textbf{0.962}\\
 & 150 & 0.026 &  0.033  & \textbf{0.791} & \textbf{0.917}\\ \midrule 
 Averaging & NA & 0.245 & 0.142 & \textbf{0.996} & \textbf{1.000} \\
\bottomrule

\end{tabular}}
\caption{Performance of different watermarking methods under the Imprint-Forgery~\cite{muller2025black}, Imprint-Removal~\cite{muller2025black}, and Averaging~\cite{yang2024steganalysis} Attacks. Note that ``G. Shad.'' denotes Gaussian Shading~\cite{yang2024gaussian}.}
\label{black_box attack}
\end{table}

\subsection{Empty Prompt Example}
\label{Empty Prompt}
\begin{figure} [h!]
	\centering
    
	\includegraphics[width=0.8\linewidth]{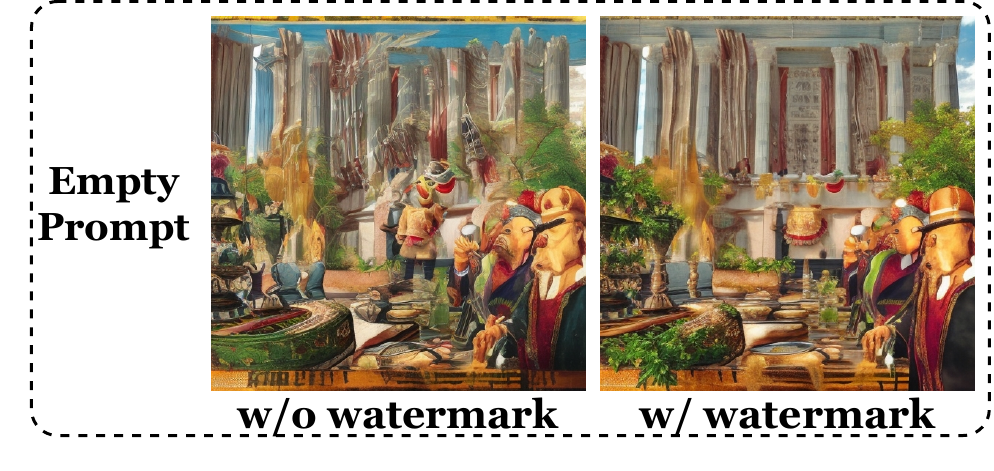}
    \vskip -0.15in
        \caption{Visualization of OptMark’s output given an empty prompt.
}
\vskip -0.05in
\label{fig:simple_pic}
\end{figure}

{We evaluate OptMark on images generated with an empty prompt. As shown in Fig.~\ref{fig:simple_pic}, this setting produces low‑quality, content‑diverse outputs. Despite these challenging conditions, OptMark maintains its full robustness.
}

\subsection{The Impact of Training Steps on Robustness}
\label{Training Steps on Robustness}
\begin{figure*}[t]
    \centering
    \includegraphics[width=\linewidth]{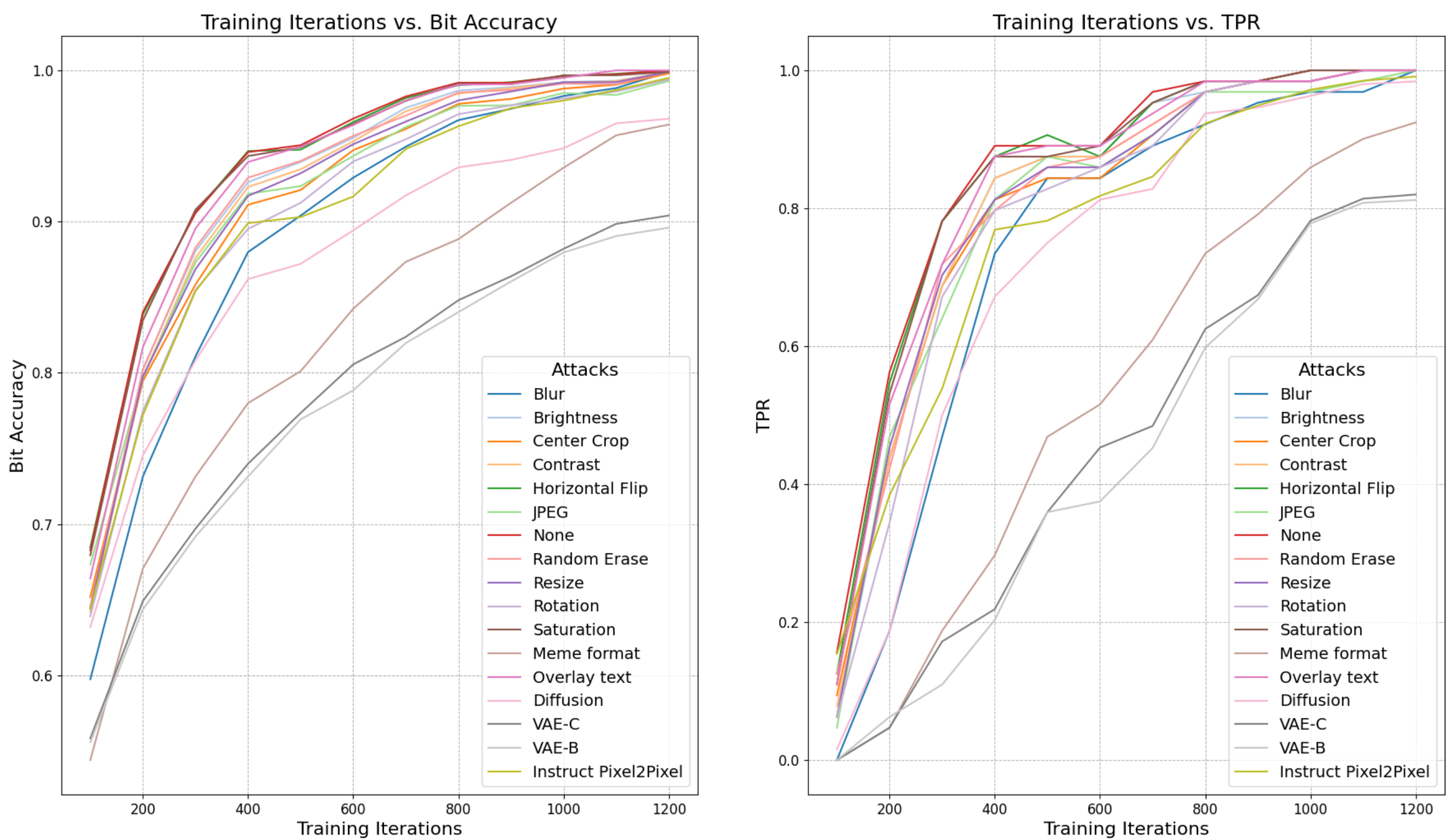}
    \caption{The relationship between the training iterations and watermark robustness.}
    \label{fig:iteration_change}
\end{figure*}

Fig.~\ref{fig:iteration_change} illustrates the relationship between training iterations and the watermark robustness of our OptMark. Larger training iterations can achieve better watermark robustness, and users can adjust the number of training iterations based on their requirements as a trade-off.

\subsection{Generality of OptMark across different diffusion samplers.
} \label{Generality}
{To demonstrate the generality of our OptMark under different diffusion samplers, we conduct additional experiments using the DPM-Solver++~\cite{lu2025dpm} sampler with 20 inference steps, with the results shown in Tab.~\ref{tab:dpm-solver++}. On a set of 1,000 images, the setting $t_d=251$  remains optimal, achieving an average multi-bit accuracy of {0.985} and a TPR of {0.973}, thereby confirming the robustness and general applicability of OptMark.}

\begin{table}[t!]
\centering
\small
\begin{tabular}{ccccccc}
\toprule
  & \multicolumn{2}{c}{\textbf{DDIM}} & \multicolumn{2}{c}{\textbf{DPM-Solver++}}   \\
 \multirow{-2}{*}{\textbf{Robustness}} & {Bit Acc.} & TPR  & {Bit Acc.} & TPR  \\ \midrule
Average   & 0.983 & 0.972 &  0.985 & 0.973 \\ \bottomrule
\end{tabular}
\vskip -0.05in

\caption{Impact of Different Diffusion Samplers on OptMark’s Robustness.
}
\vskip -0.05in
\label{tab:dpm-solver++}
\end{table}

\subsection{More Qualitative Results}
\label{Qualitative Results}
\begin{figure*} [t!]
	\centering
	\includegraphics[width=0.88\linewidth]{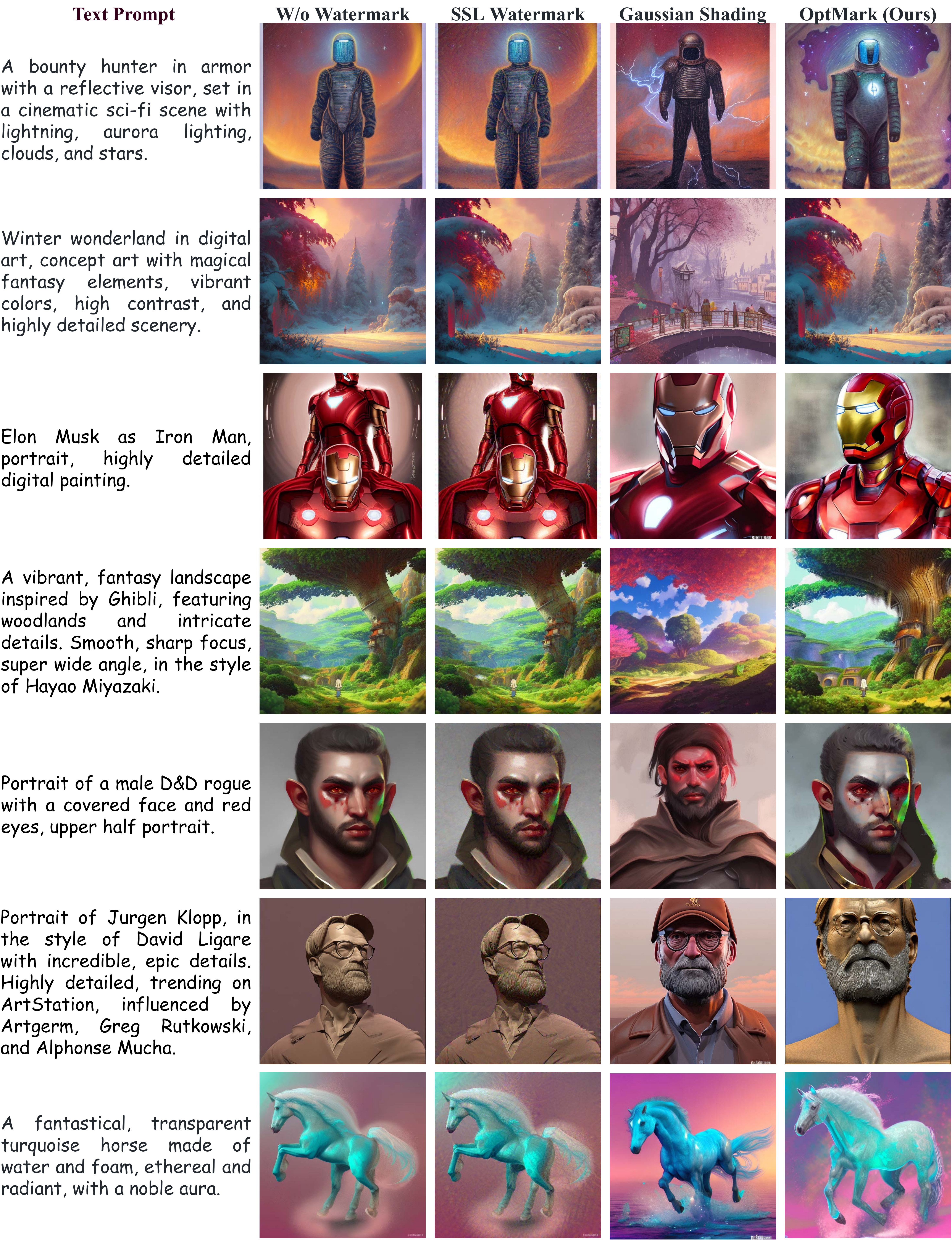}
	\caption{More qualitative comparison results between SSL Watermark~\cite{fernandez2022watermarking}, Gaussian Shading~\cite{yang2024gaussian}, and our proposed OptMark.}
	\label{fig:more_result}
\end{figure*}

More qualitative results are shown in Fig.~\ref{fig:more_result}.
\begin{table*}[t]
\centering
\small
\setlength{\tabcolsep}{1pt}
\begin{tabular}{c *{17}{c}}
\toprule
& & \multicolumn{2}{c}{DwtDct} & \multicolumn{2}{c}{DwtDctSvd} & \multicolumn{2}{c}{RivaGAN*} &  \multicolumn{2}{c}{SSL} & \multicolumn{2}{c}{S. Sign.}& \multicolumn{2}{c}{G. Shad.}& \multicolumn{2}{c}{AquaLoRA} & \multicolumn{2}{c}{\textbf{OptMark}} \\
\cmidrule(lr){3-4} \cmidrule(lr){5-6} \cmidrule(lr){7-8} \cmidrule(lr){9-10} \cmidrule(lr){11-12} \cmidrule(lr){13-14} \cmidrule(lr){15-16}  \cmidrule(lr){17-18} 
& & \multicolumn{1}{c}{Bit acc.} & \multicolumn{1}{c}{TPR}
& \multicolumn{1}{c}{Bit acc.} & \multicolumn{1}{c}{TPR}
& \multicolumn{1}{c}{Bit acc.} & \multicolumn{1}{c}{TPR}
& \multicolumn{1}{c}{Bit acc.} & \multicolumn{1}{c}{TPR}
& \multicolumn{1}{c}{Bit acc.} & \multicolumn{1}{c}{TPR}
& \multicolumn{1}{c}{Bit acc.} & \multicolumn{1}{c}{TPR}
& \multicolumn{1}{c}{Bit acc.} & \multicolumn{1}{c}{TPR}
& \multicolumn{1}{c}{Bit acc.} & \multicolumn{1}{c}{TPR} \\
\midrule
& None & 0.828 & 0.576 & \textbf{1.000} & \textbf{1.000} & 0.994 & 0.994 & \textbf{1.000} & \textbf{1.000} & 0.995 & 0.998 & \textbf{1.000} & \textbf{1.000} & 0.963 & 0.979 & \textbf{1.000} & \textbf{1.000} \\ \midrule
\multirow{4}{*}{\rotatebox{90}{\shortstack{Geomtric}}} 
&  Horizontal Flip  & \underline{0.474}& \underline{0.000} & \underline{0.438} & \underline{0.000} & \underline{0.506} & \underline{0.000} &   \textbf{1.000} & \textbf{1.000} & \underline{0.676} & \underline{0.000} & \underline{0.553} & \underline{0.000} & \underline{0.651} & \underline{0.000}& \textbf{1.000} & \textbf{1.000} \\
& Rotation (40) & \underline{0.502} & \underline{0.000} & \underline{0.471} & \underline{0.000} & \underline{0.499} & \underline{0.000} & 0.991 & 0.998 & \underline{0.621} & \underline{0.000} & \underline{0.485} & \underline{0.000} & \underline{0.478} & \underline{0.000} & \textbf{0.994} & \textbf{1.000} \\
& Resize (0.6) & \underline{0.503} & \underline{0.000} & \underline{0.476} & \underline{0.000} & 0.973 & 0.986 & 0.995 & 0.994 & 0.951 & 0.982 & \textbf{0.999} & \textbf{1.000} & 0.962 & 0.976 & \textbf{0.999} & \textbf{1.000} \\
& Crop (0.6) &\underline{0.526} &\underline{0.000} &\underline{0.488} &\underline{0.000} &0.991 &0.984 &0.997 &0.998 &0.993 &\textbf{1.000} &\underline{0.497} &\underline{0.000} &\underline{0.667} & \underline{0.106} & \textbf{0.998} & 0.998 \\ \midrule
\multirow{5}{*}{\rotatebox{90}{\shortstack{Valuemetric}}} 
&  Blur (11) & \underline{0.529} & \underline{0.000} & 0.986 & \textbf{1.000} & 0.984 & 0.974 & \textbf{0.999} & \textbf{1.000} & \underline{0.526} & \underline{0.000} & \textbf{0.999} & \textbf{1.000} & 0.960 &  0.979 & \textbf{0.999} & \textbf{1.000} \\
& Brightness (0.5) &\underline{0.489} &\underline{0.000} &\underline{0.635} &\underline{0.032} &0.976 &0.965 &\textbf{0.999} &\textbf{1.000} &0.992 &0.998 &\textbf{0.999} &\textbf{1.000} & 0.955  & 0.975 &\textbf{0.999} &\textbf{1.000} \\
& JPEG (50) &\underline{0.499} & \underline{0.000} & 0.889 &0.800 & 0.942 & 0.932 & 0.949 & 0.972 & \underline{0.640} & \underline{0.646} &\textbf{0.993} &0.984 &0.950 & 0.970 &\textbf{0.993} &\textbf{1.000} \\
& Contrast (0.5) & \underline{0.488} & \underline{0.000} &\underline{0.415} & \underline{0.032} &0.974 & 0.972 &0.999 &\textbf{1.000} &0.970 &0.980 &0.999 &\textbf{1.000}  &0.951 &0.972 &\textbf{1.000}&\textbf{1.000} \\ 
& Saturation (1.5) &\underline{0.540} &\underline{0.090} &\underline{0.580} &\underline{0.162} &0.992 &0.986 &0.999 &\textbf{1.000} &0.994 &0.996 &\textbf{1.000} &\textbf{1.000} & 0.953& 0.966&0.999 &\textbf{1.000} \\ \midrule
\multirow{4}{*}{\rotatebox{90}{\shortstack{Editing}}} 
& Meme Format  & {0.796}  & \underline{0.453} &0.852  & 0.666  & 0.974  & 0.982  &\textbf{0.981}  & \textbf{1.000}  & \underline{0.579}  & \underline{0.016}&\underline{0.481} &\underline{0.000} & \underline{0.643}  &\underline{0.000}   &0.964 & 0.925 \\
& Random Erase (0.1)  &0.774 &\underline{0.422} &0.998 &1.000 &0.993 &0.996 &\textbf{0.999} &\textbf{1.000} &\underline{0.577} &\underline{0.000} &\textbf{0.999} &\textbf{1.000}&0.929 & 0.945&\textbf{0.999} &\textbf{1.000} \\
& Text Overlay  & 0.828  & 0.576  &\textbf{1.000}  &\textbf{1.000}  & 0.991 & 0.988   &\textbf{1.000}  & \textbf{1.000}  & 0.991 & 0.996  &\textbf{1.000} &\textbf{1.000} & 0.950 & 0.975 &\textbf{1.000}  &\textbf{1.000}  \\
& InstructPix2Pix  &\underline{0.478}  & \underline{0.000}  &\underline{0.496}  & \underline{0.016}  & \underline{0.699} &\underline{0.123}  &\underline{0.708}  & \underline{0.000} & \underline{0.542}  & \underline{0.000}  &\textbf{0.997} &\textbf{1.000} &0.911 &0.886 &0.995 & 0.991 \\  \midrule
\multirow{3}{*}{\rotatebox{90}{\shortstack{Regen-\\ eration}}} 
& VAE-B (3) &\underline{0.493} & \underline{0.000} & \underline{0.612} & \underline{0.002} &\underline{0.567} &\underline{0.002} &\underline{0.626} &\underline{0.008} &\underline{0.639} &\underline{0.016} &\textbf{0.980} &\textbf{0.937}& 0.936& 0.964 &0.896 &0.812 \\
& VAE-C (3) &\underline{0.493} &\underline{0.000} &\underline{0.602} &\underline{0.016} & \underline{0.553}& \underline{0.000} & \underline{0.579} & \underline{0.004} & \underline{0.651}& \underline{0.018} & \textbf{0.981} &\textbf{0.953} &0.940 &0.970&0.904 & 0.820 \\
& Diffusion (60) &\underline{0.495} &\underline{0.000} &\underline{0.602} &\underline{0.048} &\underline{0.590} &\underline{0.004} &\underline{0.582} &\underline{0.002} &\underline{0.527} &\underline{0.000} &\textbf{0.997} &\textbf{0.984}& 0.915& 0.930 &0.968 &\textbf{0.984} \\ \midrule

& \textbf{Average-attack} & \underline{0.573} & \underline{0.125} & \underline{0.679} & \underline{0.340} & 0.835 & 0.641 & 0.906 & 0.763 & 0.757 & 0.509 & {0.880} & {0.756}& 0.866 & 0.741 & \textbf{0.983} & \textbf{0.972}  \\\bottomrule
\end{tabular}
\vskip -0.05in
\caption{Full detecting results of different watermarking methods under various attacks on DiffusionDB~\cite{GustavostaStableDiffusionPromptsDatasets2023}.  ``*'' indicates that Gaussian Shading~\cite{yang2024gaussian} and RivaGAN~\cite{zhang2019robust} can embed 64-bit and 32-bit hidden messages due to the method constraint, whereas all other methods are compared under the condition of embedding 48-bit messages. ``Average-attack'' indicates calculating the average score across cases under sixteen different attacks and the no-attack (``None''). The \underline{underline} indicates poor robust performance with Bit Acc. $< 0.75$ and TPR $< 0.5$. Note that ``S. Sign." and ``G. Shad.'' denote Stable Signature~\cite{fernandez2023stable} and Gaussian Shading~\cite{yang2024gaussian}, respectively.} 
\vskip -0.05in

\label{table:detal_compare}
\end{table*}

\section{Limitations}
\label{Limit}
{Although our OptMark demonstrates strong robustness against various types of attacks, its performance under regeneration attacks is slightly inferior to that of other semantic-level watermarking methods, such as Gaussian Shading. We believe the performance difference lies in the extraction methods: Gaussian Shading uses inversion to extract the watermark from the initial noise space, while OptMark uses the DINO network. We suspect that the watermark recovered from the inverted initial noise is inherently more robust to regeneration attacks than from the DINO latent space. To address this, we plan to explore alternative watermark extractors that offer more resilient extraction spaces—such as the denoising UNet used in diffusion models in the future.}

\end{document}